\documentclass{aa}  

\usepackage{graphicx}
\usepackage{txfonts}
\usepackage[colorlinks=true, linktocpage, linkcolor={blue!60!black}, citecolor={blue!60!black}, urlcolor={blue!60!black}]{hyperref}
\usepackage{xcolor}
\usepackage{multicol}

\begin{document} 
   \title{A theoretical investigation of far-infrared fine structure lines at $z>6$ and of the origin of the {\rm [OIII]}$_{88\mu m}$/{\rm [CII]}$_{158\mu m}$ enhancement}
   
   \titlerunning{A theoretical investigation of FIR fine structure lines at $z>6$}

   \author{C. T. Nyhagen
          \inst{1,2}
          \and A. Schimek\inst{1}
          \and C. Cicone\inst{1}
          \and D. Decataldo \inst{1}
          \and S. Shen \inst{1}
          }

   \institute{Institute of Theoretical Astrophysics, University of Oslo, PO Box 1029, Blindern 0315, Oslo, Norway \\
         \and
             Lund Observatory, Division of Astrophysics, Department of Physics, Lund University, SE-221 00 Lund, Sweden\\
            \email{camilla.nyhagen@fysik.lu.se}
             }

   \date{Received 23 October 2024; accepted 17 July 2025}

  \abstract{
  The far-infrared (FIR) fine structure lines [CII]$_{158\mu m}$, [OIII]$_{88\mu m}$, [NII]$_{122\mu m}$, and [NIII]$_{57\mu m}$ are excellent tools for probing the physical conditions of the interstellar medium (ISM). The [OIII]$_{88\mu m}$/[CII]$_{158\mu m}$ and [OIII]$_{88\mu m}$/[NII]$_{122\mu m}$ luminosity ratios have shown to be promising tracers of the ionisation state and gas-phase metallicity of the ISM. Observations of galaxies at redshift $z > 6$ show unusually high [OIII]$_{88\mu m}$/[CII]$_{158\mu m}$ luminosity ratios compared to local sources. The origin of the enhanced ratios has been investigated in the literature with different theoretical modelling approaches. However, no model has to date successfully managed to match the observed emission from both [OIII]$_{88\mu m}$ and [CII]$_{158\mu m}$, as well as their ratio. For this study we used \textsc{Cloudy} to model the [CII]$_{158\mu m}$, [OIII]$_{88\mu m}$, [NII]$_{122\mu m}$, and [NIII]$_{57\mu m}$ emission lines of \textsc{Ponos}, a high-resolution ($m_{\mathrm{gas}} = 883.4\, M_{\odot}$) cosmological zoom-in simulation of a galaxy at redshift $z = 6.5$, which is post-processed using \textsc{kramses-rt}. We modify carbon, nitrogen, and oxygen abundances in our \textsc{Cloudy} models to obtain C/O and N/O abundance ratios respectively lower and higher than solar, more in line with recent high-z observational constraints. We find [OIII]$_{88\mu m}$/[CII]$_{158\mu m}$ luminosity ratios that are a factor of $\sim 5$ higher compared to models assuming solar abundances. Additionally, we find an overall better agreement of the simulation with high-z observational constraints of the [CII]$_{158\mu m}$-SFR and [OIII]$_{88\mu m}$-SFR relations. This shows that a lower C/O abundance ratio is essential to reproduce the enhanced [OIII]$_{88\mu m}$/[CII]$_{158\mu m}$ luminosity ratios observed at $z > 6$. By assuming a super-solar N/O ratio, motivated by recent $z > 6$ JWST observations, our models yield an [OIII]$_{88\mu m}$/[NII]$_{122\mu m}$ ratio of $1.3$, which, according to current theoretical models, would be more appropriate for a galaxy with a lower ionisation parameter than the one we estimated for \textsc{Ponos}. Most current simulations adopt solar abundance patterns that are not adequate for recently observed high-z predictions. Our results showcase the importance of theoretical modelling efforts, coupled with high-resolution zoom-in simulations, and with parallel multi-tracer observations to understand the physical and chemical conditions of the ISM at $z > 6$.}

   \keywords{galaxies: evolution -- galaxies: high-redshift -- galaxies: ISM -- methods: numerical -- ISM: lines and bands -- ISM: abundances}

   \maketitle

\section{Introduction}

\begin{figure*}[ht!]
    \centering
    \includegraphics[width=0.7\textwidth]{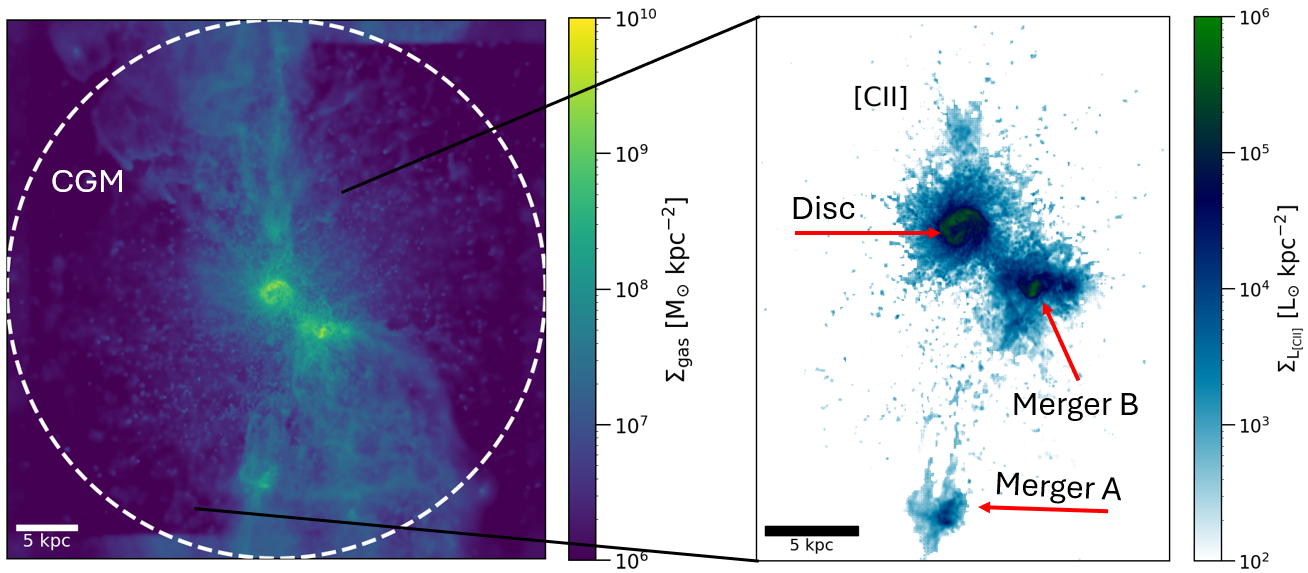}
    \caption{Gas surface density map of the \textsc{Ponos} simulation (left panel). 
    The dashed circle corresponds to the virial radius, $R_{\mathrm{vir}}$.
    The right panel shows the $\mathrm{[CII]}$ line surface density map obtained using the fiducial model of \cite{Schimek23, Schimek24}, which assumes solar elemental abundances, zoomed into the galaxy disc and the merging companions.}  
    \label{fig:overview}
\end{figure*}

With the formation of the first stars and galaxies, the intergalactic medium (IGM) transitioned from neutral to ionised during the Epoch of Reionisation (EoR) \citep{Robertson22}, which has been empirically shown to have ended at $z \approx 6$ \citep{Fan02,Stark10, Robertson15}. At $z > 6$, galaxies are less dynamically settled as they experience cold mode accretion \citep{Dekel09} and frequent interactions \citep{Lacey1993,Hopkins10, Somerville15, CRISTAL_Posses24}. 
The gas-phase metallicity in the interstellar medium (ISM) is expected to be low at $z > 6$, due to core-collapse supernovae (cc-SNe) being the main enrichment mechanism on short timescales \citep{Maiolino_Mannucci19}.
By pushing the frontiers of high-z observations, the James Webb Space Telescope (JWST) has spectroscopically confirmed luminous galaxies out to $z \sim 14$ \citep{Carniani24, Schouws24, Naidu25}, enabling us to study galaxies at the beginning of the EoR. Understanding the properties of the ISM at $z > 6$ is crucial to gaining insight into the early stages of galaxy formation and evolution. 

At $z > 6$, commonly employed ISM gas tracers producing rest-frame lines in the ultraviolet and optical bands are redshifted to wavelengths inaccessible from Earth. However, far-infrared (FIR) emission lines from atomic and ionised gas are redshifted into the sub-millimetre (sub-mm) atmospheric windows. They are thus observable with ground-based facilities placed on high-altitude sites, such as the Atacama Large Millimeter/submillimeter Array (ALMA) and, in the future, the Atacama Large Aperture Submillimeter Telescope (AtLAST).\footnote{under development, consult: \href{https://www.atlast.uio.no/}{atlast-telescope.org}}
The advantage of FIR lines is that they are bright tracers of the cold and warm ISM where star formation is occurring \citep{deLooze14, Herrera15}, in addition to being less affected by dust-extinction compared to optical lines \citep{Brauher08, Ferkinhoff10}.
Most high-z sub-mm observations target the emission lines from singly ionised carbon (C$^+$), $\mathrm{[CII]_{158\mu m}}$ (hereafter $\mathrm{[CII]}$), and from doubly ionised oxygen (O$^{++}$), $\mathrm{[OIII]_{88\mu m}}$ (hereafter $\mathrm{[OIII]}$).
$\mathrm{[CII]}$ is one of the main coolants of the ISM \citep{Tielens85, Wolfire22}, making up   0.1~-~1\% of the total FIR emission of a star-forming galaxy \citep{Stacey91}. Due to its ionisation potential ($\mathrm{11.26~eV}$) lower than hydrogen ($\mathrm{13.6~eV}$) \citep{Cormier15}, C$^{+}$ can reside co-spatially with neutral atomic, molecular, and ionised gas \citep{Olsen15, Olsen17, Lagache18, Leung20}. Most of the [CII] emission is believed to originate from photodissociation regions (PDRs) \citep{Stacey10}. 
$\mathrm{[OIII]}$ is a tracer of the warm-hot ionised gas (T~$\sim$ 10$^{4}$ - 5~$\times$~10$^{5}$~K), and of HII regions \citep{Cormier15}. 
Due to the high ionisation potential of O$^{++}$ (35.1 eV), [OIII] probes radiation from hot young stars and is thus a good extinction-free tracer of star formation \citep{Ferkinhoff10}. 
The [CII] and [OIII] emission lines have been investigated in $z > 6$ observations (e.g. \citealt{Kanekar13, Ouchi13, Capak15, Maiolino15, Inoue16, Laporte_2017, Hashimoto18, Tamura19, Schaerer20, Vanderhoof22, CRISTAL_Posses24, Schouws24, Schouws25}); however, the number of galaxies where both [CII] and [OIII] are detected is still limited (e.g. \citealt{Inoue16,Carniani17, Hashimoto19, Harikane20, Witstok22,Fujimoto22, Ren23, Algera23, Kumari24, Schouws25}). Both $\mathrm{[CII]}$ and $\mathrm{[OIII]}$ have been empirically shown to correlate with the star formation rate (SFR) of galaxies \citep{deLooze14,Herrera15}.
Additionally, since the [OIII] and [CII] emission lines trace different gas phases, we can use their luminosity ratios to study the ionisation state of the ISM \citep{Harikane20}.
In local galaxies, the $\mathrm{[OIII]/[CII]}$ luminosity ratio is observed to be $\lesssim 1$, while for galaxies at $z > 6$ with comparable SFR, this value is significantly higher with $\mathrm{[OIII]/[CII]}$ values $\sim 1-20$ \citep{Harikane20}. 
The origin of the enhanced high-z $\mathrm{[OIII]/[CII]}$ ratios is still debated. Several contributing factors have been proposed, such as lower PDR covering fractions, higher ionisation parameters, and lower gas-phase metallicities \citep{Harikane20, Arata20, Lupi2020, Vallini21, Katz22}.

The $\mathrm{[NII]_{122\mu m}}$ line (hereafter $\mathrm{[NII]}$) from singly ionised nitrogen $\mathrm{N^+}$, and $\mathrm{[NIII]_{57\mu m}}$ (hereafter $\mathrm{[NIII]}$), from doubly ionised nitrogen $\mathrm{N^{++}}$, are both tracers of ionised gas \citep{Cormier15}, with ionisation potentials of $\mathrm{14.53~eV}$ and $\mathrm{29.6~eV}$ respectively.
Together with [CII] and [OIII], [NII] and [NIII] can be used to study the ionisation parameter and gas-phase metallicity of the ISM \citep{Nagao11,Pereira17, Rigo18, Schimek24}. 
The current number of observational constraints for the [NII] and [NIII] lines is limited. The [NII] line has only been observed in a few non-quasar galaxies at $z > 6$, with the detection of $\mathrm{[NII]_{122\mu m}}$\citep{Killi23} and of $\mathrm{[NII]_{205\mu m}}$ \citep{tadaki22, kolupuri25, fudamoto25}, while the [NIII] line has yet to be detected at such high redshifts.

Theoretical efforts can help us place the properties of the high-z ISM into the broader context of galaxy formation and evolution. Recent studies include semi-analytical models (e.g. \citealt{Lagache18, Popping19}), cosmological zoom-in simulations focusing on small galaxy samples ($\lesssim 10-30$ galaxies; e.g. \citealt{Vallini15,Vallini18, Vallini21, Katz19, Olsen17, Pallottini19, Lupi2020, Arata20}), and simulations with large samples of galaxies (\citealt{Katz22, Pallottini22, Padilla23, Nakazato25}). Despite the variety of theoretical modelling approaches, no model has been successful at simultaneously reproducing the high [OIII]/[CII] ratios observed at $z > 6$ and the [CII] and [OIII] luminosity values consistent with the high-z [CII]-SFR and [OIII]-SFR empirical relations. 
In contrast to [OIII] and [CII], the [NII] and [NIII] lines at $z \gtrsim 6$ have only been investigated in a few theoretical studies \citep{Padilla23, Schimek24}, resulting in the [NII]-SFR and [NIII]-SFR relations being poorly constrained at high z. 

Most current theoretical models assume solar abundance patterns when accounting for the ISM chemistry, as it is computationally expensive to trace the element abundances directly for each cell \citep{Olsen17, Vallini18, Pallottini19,Pallottini22, Padilla23}. However, recent observational constraints from JWST have revealed that the chemistry at $z > 6$ could be vastly different from the local ISM, with observations of sub-solar C/O abundance ratios and super-solar N/O abundance ratios \citep{Jones23, Isobe23, Cameron23}, making the assumption of solar values unrealistic when modelling the high-z ISM. In previous theoretical works, the assumptions of ISM enrichment patterns at high redshifts are based on supernovae (SNe) and stellar yields. With C/O and N/O estimates from the same $z > 6$ sources now becoming available, we can implement the observed estimates in our theoretical models and compare the results to the observed high-z galaxy properties. 

For this work, our aim was to make the modelling of the [CII], [OIII], [NII], and [NIII] lines more representative of high-z galaxies that are mainly enriched by cc-SNe. To this end, we adopted the fiducial radiative transfer (RT) approach from \cite{Schimek23} (also used in the follow-up work, \cite{Schimek24}), who modelled several FIR and sub-mm emission lines of the high-resolution (spatial resolution $<4$~pc) \textsc{Ponos} simulation at $z = 6.5$ \citep{Ponos}. We changed the carbon, nitrogen, and oxygen abundances in our \textsc{Cloudy} models to better match the C/O and N/O abundance ratios observed in $z > 6$ galaxies and/or local analogues. This approach was inspired by \cite{Katz22}, who assumed cc-SNe yields \citep{Nomoto06} in their \textsc{Cloudy} models and applied them to their simulated sample of $z = 4-10$ galaxies. We studied how the change in abundance ratios affects the resulting $\mathrm{[OIII]/[CII]}$ and $\mathrm{[OIII]/[NII]}$ ratios and the individual line luminosity values.  
This paper is structured as follows: in Sect.~\ref{sec:simulation} we briefly introduce the properties of the fiducial model from \cite{Schimek23} and describe our methods for modelling the emission lines. In Sect.~\ref{sec:abund} we present our three models based on observational constraints of elemental abundance ratios. Our results are shown and discussed in Sect.~\ref{sec:results} and Sect.~\ref{sec:discussion}. Our conclusions summarising our most important findings are presented in Sect.~\ref{sec:conclusion}.

\section{Simulation and modelling}\label{sec:simulation}
\subsection{The \textsc{Ponos} simulation}

\cite{Schimek23} modelled the FIR and sub-mm line emission arising from the ISM and circumgalactic medium (CGM) of the \textsc{Ponos} simulation \citep{Ponos}: a high-resolution cosmological zoom-in simulation of a typical star-forming galaxy halo at redshift $z = 6.5$, which is the progenitor of a massive galaxy ($M_\mathrm{{vir}}(z = 0) = 1.2\times 10^{13}$ $M_{\odot}$) at z = 0. The \textsc{Ponos} simulation is run with the \textsc{Gasoline} smoothed particle hydrodynamics (SPH) code \citep{Wadsley04}, and uses $\mathrm{WMAP\,7/9}$ cosmology ($\mathrm{\Omega_{m,0} = 0.272}$, $\mathrm{\Omega_{\Lambda, 0} = 0.728}$, $\mathrm{\Omega_{b,0} = 0.0455}$, $\sigma_8 = 0.807$, $n_\mathrm{s} = 0.961$, and $H_0 = 70.2 \, \mathrm{km \, s}^{-1} \mathrm{Mpc}^{-1}$; \citealt{Komatsu11, Hinshaw13}). The simulation has a gas mass resolution of $m_{\mathrm{gas}} = 883.4$ $M_{\odot}$, a stellar particle mass of $m_{*}$ = 0.4$\cdot m_{\mathrm{gas}} = 353.4\,M_{\odot}$,
and a minimum smoothing length of $\sim3.6 \, \mathrm{pc}$.
The virial mass of \textsc{Ponos} at $z = 6.5$ is $M_{\mathrm{vir}}(z = 6.5) = 1.22\times 10^{11}\,M_{\odot}$.
\textsc{Ponos} has a stellar mass $M_{*} = 2\times 10^9\,M_{\odot}$, a virial radius of $R_{\mathrm{vir}} = 21.18$ kpc and a star formation rate (SFR) = $20\,M_{\odot}$ yr$^{-1}$. Similarly to \cite{Schimek23}, we define the galaxy halo to be confined within $R_{\mathrm{vir}}$ and divided it further into separate components: main disc, merging companions, and CGM (see \cite{Schimek23} for further details). In this work, we focus on the global line emission from the simulation and also the emission from the brightest ISM components (main disc and mergers), hence we do not explore the variations between the ISM and the CGM.
An overview of the simulation is shown in Fig.~\ref{fig:overview},
which also reports the $\mathrm{[CII]}$ map obtained by using the same fiducial model as in \cite{Schimek23, Schimek24}. We have used \textsc{Pynbody} \citep{Pynbody} for the analysis of the simulation. 

The fiducial model from \cite{Schimek23} uses radiative transfer post-processing with \textsc{kramses-rt} \citep{Pallottini19, Decataldo19, Decataldo20}, to model the propagation of radiation from all emitting sources. \textsc{kramses-rt} is a customised version of the adaptive mesh refinement (AMR) code \textsc{ramses-rt} \citep{Ramses, RamsesRT} featuring a non-equilibrium chemical network generated by \textsc{krome} \citep{KROME}. For the post-processing, the SPH simulation was converted into an AMR grid (see \citealt{Schimek23} for details), with eight different refinement levels, resulting in grid sizes ranging from $\mathrm{1.52~pc}$ for the high-density regions (higher resolution) to $\mathrm{195.3~pc}$ for the low-density regions (lower resolution). Since the dynamical evolution is turned off in the RT post-processing, there is no mixing of the hot and cold gas. This can be seen in the bottom panels of Fig. 8 in \cite{Schimek23}, where they show the density-temperature phase diagrams for their fiducial RT model. \cite{Schimek23} modelled the emission lines of \textsc{Ponos} using the 1D photoionisation code \textsc{Cloudy} \citep{Cloudy}, by first creating multidimensional \textsc{Cloudy} grids and then interpolating to the values in the simulation cell.
As the \textsc{Ponos} simulation does not trace elements directly for each particle, \cite{Schimek23, Schimek24} assumed solar abundances using the default \textsc{Cloudy} setup \citep{Grevesse98, Holweger01,Allende_Prieto_2001,Allende_Prieto_2002}. 

\subsection{Post-processing with \textsc{Cloudy}} \label{sec:cloudy-pp}

\begin{table}[]
\centering
\caption{\textsc{Cloudy} grid parameters.}
\begin{tabular}{lcccc} 
    \hline
    \hline
    Parameter & Min & Max & Steps & Extra values\\
    \hline
    log($T$) & $2.0$ & $6.0$ & $0.5$ & $5.2, 5.7$\\
    log($n_\mathrm{H}$)  & $-3.5$ & $4.0$ & $0.5$ & none\\
    log$(Z)$ & $-3.0$ & $0.6$ & $0.6$ & none\\
    log($G_0$) & $-4.9$ &$4.2$ & $0.6$ & none \\
    \hline
\end{tabular}
\tablefoot{The first column shows the four parameters: temperature log($T$) [K], density log($n_\mathrm{H}$) [cm$^{-3}$], gas-phase metallicity log$(Z)~[Z_{\odot}]$, and radiation field log($G_0$) [$1.6\times10^{-3}$ erg cm$^{-2}$ s$^{-1}$]. The second, third, and fourth columns show the minimum and maximum values of each parameter and the step size. The rightmost column shows additional values that have been added that are not part of the original parameter range.}
\label{tab:Cloudy}
\end{table}

To model the emission lines of \textsc{Ponos}, following the setup in \cite{Schimek23}, we created a multidimensional grid consisting of individual \textsc{Cloudy} models, using slab geometry. This type of grid modelling has been used in previous theoretical studies (e.g. \citealt{Olsen15,Olsen18,Pallottini19,Pallottini22,Katz19,Katz22,Lupi2020}).
Our grid consists of four parameters: temperature log($T$)~[K], density log($n_\mathrm{H}$)~[cm$^{-3}$], metallicity log$(Z)~[Z_{\odot}]$, and radiation field log($G_0$) [$1.6\times10^{-3}$ erg cm$^{-2}$ s$^{-1}$], where their maximum and minimum values are taken directly from the \textsc{kramses-rt} post-processed simulation. 
When referring to the metallicity, we refer to the gas-phase metallicity, unless otherwise stated.
Due to the post-processing with \textsc{kramses-rt}, we kept the temperature in the \textsc{Cloudy} models constant, similar to the fixed-temperature models in \cite{Lupi2020}. Note that there are inconsistencies when combining \textsc{kramses-rt} and \textsc{Cloudy} in the post-processing, as \textsc{kramses-rt} solves non-equilibrium photochemistry and \textsc{Cloudy} assumes photoionisation equilibrium. However, when comparing different FIR modelling approaches for their simulated $z = 6$ galaxy, \cite{Lupi2020} found good agreement between their fixed-temperature \textsc{Cloudy} models and the predictions from the non-equilibrium chemistry using \textsc{krome} only. 

The size range and abundance of dust were set to follow typical values for the local ISM, which reproduces the observed extinction properties of our galaxy, where we set the dust abundance and depletion of heavier elements to scale linearly with the $Z$ parameter. With \textsc{Ponos} being a galaxy at $ z = 6.5$, this could lead to an overestimation of the dust abundance where the metallicity is high; however, $< 8\%$ of the gas have super-solar metallicities. Since we only consider FIR emission lines in this work, we do not expect dust to have a significant effect in terms of dust attenuation of the emission lines, but we acknowledge that it can affect the amount of metals that emit the FIR emission lines. At $z = 6.5$, the CMB temperature is $\mathrm{20~K}$, which was included in the modelling. To compute the $\mathrm{[CII]}$, $\mathrm{[OIII]}$, $\mathrm{[NII]}$ and $\mathrm{[NIII]}$ line luminosities, considering the AMR structure in \textsc{kramses-rt}, we integrated the emission from \textsc{Cloudy} along the depth corresponding to the length of the AMR grid-cells and multiply the integral with the area of the cell. 
We refer to \cite{Schimek23} for a more in-depth description of their fiducial model and the post-processing with \textsc{kramses-rt} and \textsc{Cloudy}. 

\subsection{Solar model with coarser grid}\label{sec:solar_coarser}
The fiducial model from \cite{Schimek23, Schimek24} has a grid consisting of $42560$ independent \textsc{Cloudy} models. Since this is computationally expensive to run, we cut down the computational cost by reducing the number of grid points, allowing us to test multiple models. To make the grid coarser, we increased the step size in the temperature and density ranges, resulting in a total of $19712$ independent \textsc{Cloudy} models. We list the final grid values for our coarser model in Table~\ref{tab:Cloudy}, where we added two additional high-temperature values (log($T$) = $5.2, 5.7$), hence changing the spacing of the original grid. This was done for more accurate $\mathrm{[OIII]}$ treatment, as $\mathrm{[OIII]}$ is more sensitive to higher gas temperatures. Such high temperatures are produced by supernovae feedback, which in \textsc{Ponos} is modelled following \cite{Stinson06}, where the cooling of hot and cold gas is delayed. However, the \textsc{kramses-rt} post-processing allows gas with temperatures $T \leq 10^6$ K to cool, while the cooling of warmer gas is artificially suppressed to take into account shock heating by stellar mechanical feedback, which is not in action without dynamical evolution. The temperatures included in our \textsc{Cloudy} models, $T = 10^2 - 10^6~\rm K$, are therefore not affected by the artificial cooling suppression \citep{Stinson06}.
To test the consistency with \cite{Schimek23}, we reproduced their fiducial model with solar abundance ratios with the coarser grid. 
We find our coarser model to be overall consistent with \cite{Schimek23, Schimek24}, with a change of $\lesssim 8\%$ in total [CII], [OIII] and [NII] luminosities compared to \cite{Schimek23, Schimek24}. The [NIII] luminosity is the most enhanced one compared to \cite{Schimek24}, with a $50\%$ higher total luminosity. This is due to our change in the temperature grid range compared to \cite{Schimek23}, where we add two additional temperature values (log($T$) = $5.7, 6.0$). Hereafter we refer to this model as the Solar model, and use the same \textsc{Cloudy} setup and grid values for the other non-solar models described in Sect.~\ref{sec:abund}.

\section{Changing the elemental abundances}\label{sec:abund}

\begin{table}
\centering
\caption{$\mathrm{C/O}$ and $\mathrm{N/O}$ ratios used for the emission line modelling.}            
\label{tab:abund}     
\centering                        
\begin{tabular}{llcccc}       
\hline\hline                
Model & Ref. & $\mathrm{log(O/H)+12}$ & $\mathrm{log(C/O)}$ & $\mathrm{log(N/O)}$\\    
\hline                   
   Solar & \textsc{Cloudy} 
   &  $8.69 $& $-0.30$ & $-0.76$\\
   High-z & J23, I23
   & $7.39$ & $-1.01$ & $-0.40$ \\
   Dwarfs & B19 & $7.85$& $-0.71$ &$-1.46$ \\
   Halo & N14& $8.02$ & $-0.65$ &none\\ 
\hline                                  
\end{tabular}
\tablefoot{The first and second columns show the models and references used (\textsc{Cloudy}: default solar abundances, J23: \cite{Jones23}, I23: \cite{Isobe23}, B19: \cite{Berg19}, N14: \cite{Nissen14}). The third, fourth, and fifth columns show the oxygen abundance, the $\mathrm{C/O}$ ratio, and the $\mathrm{N/O}$ ratio.}
\end{table}

\begin{figure*}[ht]
    \centering
    \includegraphics[width=0.7\textwidth]{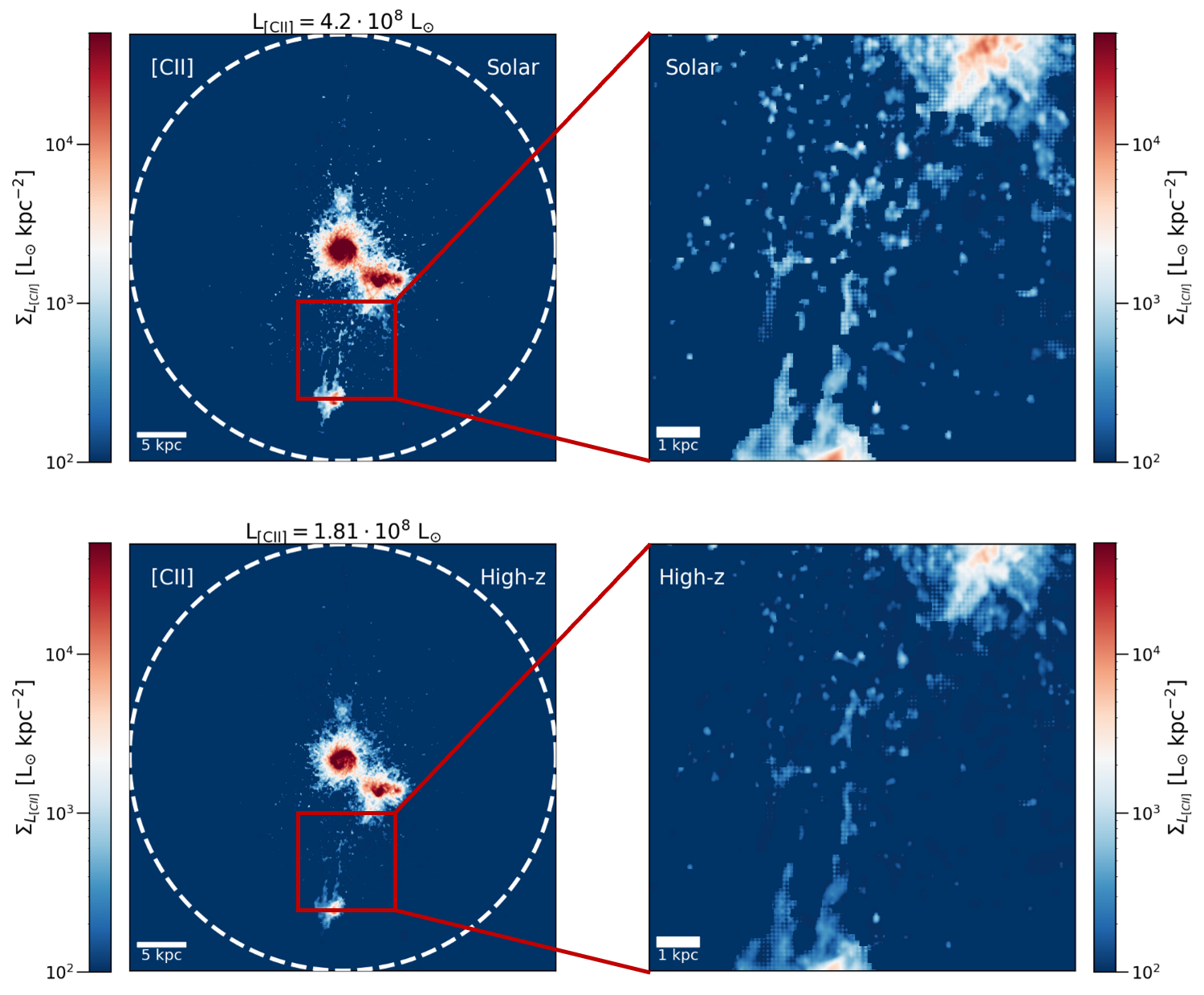}
    \caption{$\mathrm{[CII]}$ line emission maps obtained with the Solar model (top panels) and the High-z model (bottom panels). The right panels are zoomed-in images of the tidal tail between merger B and merger A. The white dashed circles mark the extent of $R_{\mathrm{vir}}$ of \textsc{Ponos}. The line emission maps obtained with the Dwarfs and Halo models are shown in Appendix~\ref{app:dwarfs}.}   
    \label{fig:CII_maps}
\end{figure*}

\begin{figure*}[ht]
    \centering
    \includegraphics[width=0.7\textwidth]{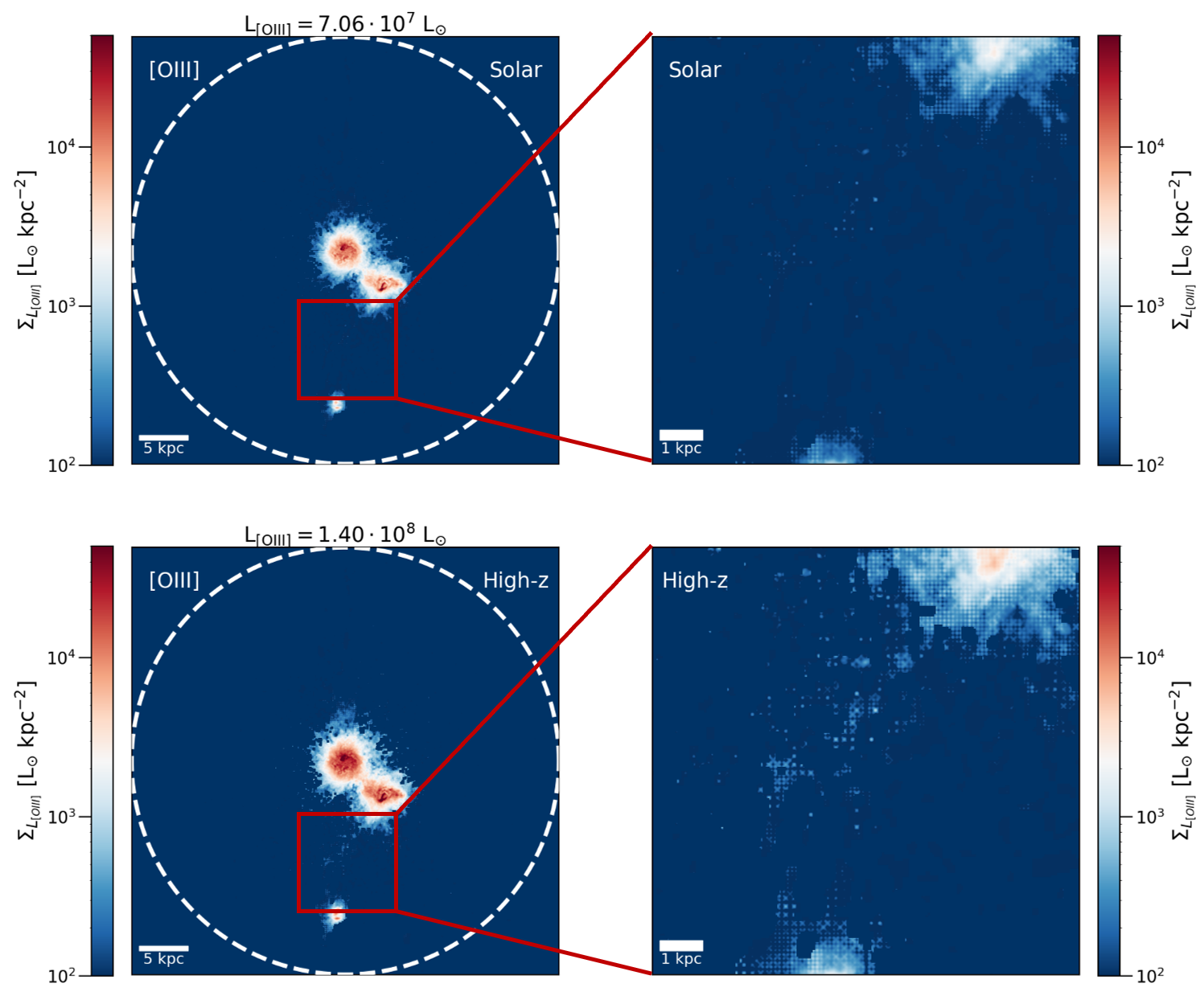}
    \caption{$\mathrm{[OIII]}$ line emission maps obtained with the Solar model (top panels) and the High-z model (bottom panels). The right panels are zoomed-in images of  the tidal tail between merger B and merger A. The white dashed circles mark the extent of $R_{\mathrm{vir}}$ of \textsc{Ponos}. The line emission maps obtained with the Dwarfs and Halo models are shown in Appendix~\ref{app:dwarfs}.}   
    \label{fig:OIII_maps}
\end{figure*}

\begin{figure*}[ht]
    \centering
    \includegraphics[width=0.7\textwidth]{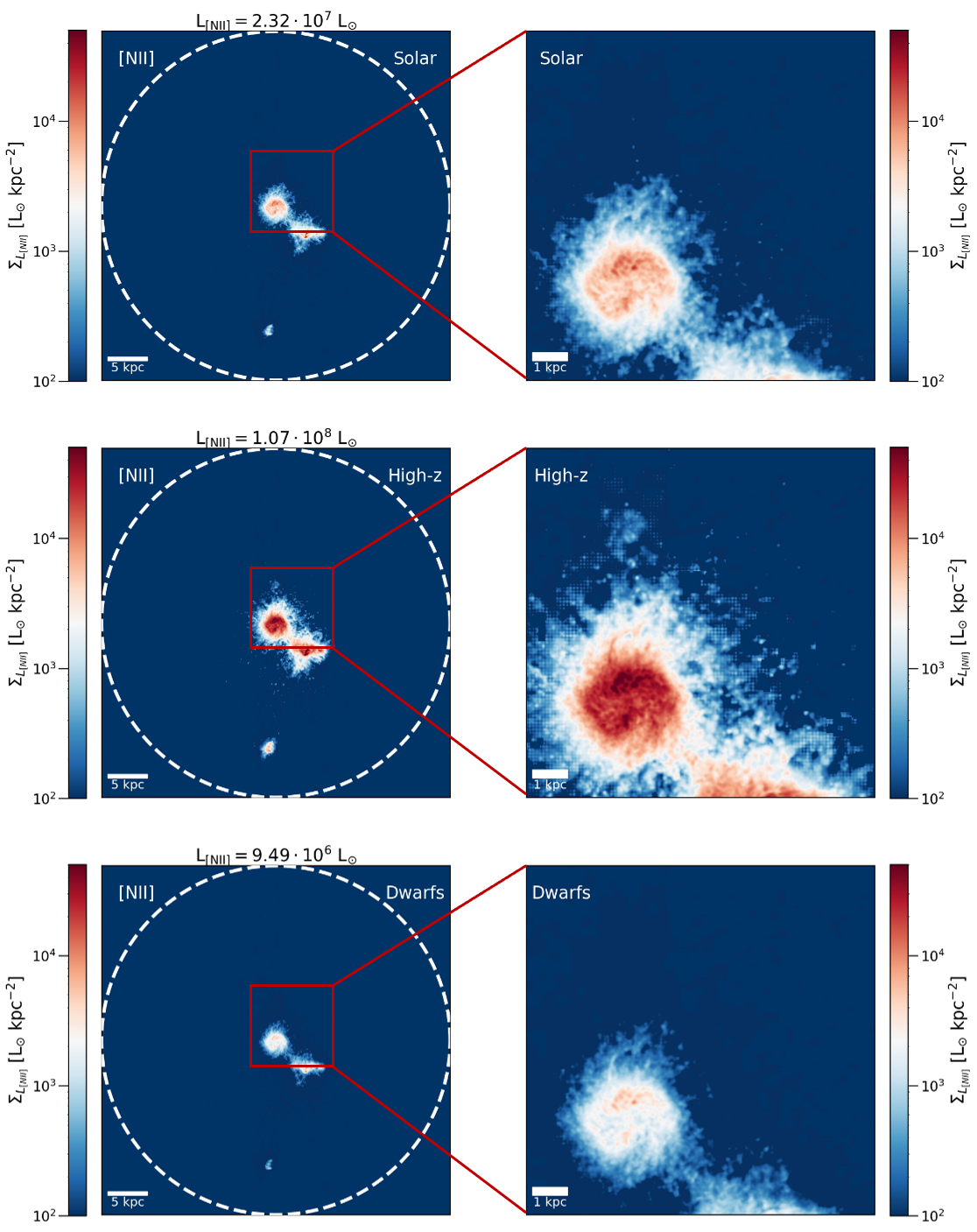}
    \caption{$\mathrm{[NII]}$ line emission maps obtained with the Solar model (top panels), the High-z model (middle panels), and the Dwarfs model (bottom panels). The right panels are  zoomed-in images of the main disc and surrounding area. The white dashed circles mark the extent of $R_{\mathrm{vir}}$ of \textsc{Ponos}.}   
    \label{fig:NII_maps}
\end{figure*}

\subsection{Observational constraints of the C/O and N/O abundance ratios}

One of the proposed contributing factors to the observed high [OIII]/[CII] luminosity ratio at $z > 6$ is a lower C/O abundance ratio \citep{Harikane20, Arata20,Katz22}.
In this work, we define abundance as the element abundance with respect to hydrogen, unless otherwise stated. Oxygen and other $\alpha$-elements (i.e. Mg, Si, S, and Ca) are produced by massive stars ($M \gtrsim 8\, M_{\odot}$), and returned to the ISM through cc-SNe \citep{Timmes95,Kobayashi06,Kobayashi20}, which have relatively short timescales of $\sim 1-10~\mathrm{Myr}$ \citep{Kobayashi06}. Carbon is also produced by these massive stars, but a significant contribution to the carbon abundance is due to winds from low-mass ($M \sim 1-4~M_{\odot}$) asymptotic giant branch (AGB) stars, which have longer lifetimes of $\mathrm{\gtrsim 100~Myr}$ \citep{Nomoto13, Kobayashi11,Kobayashi20}. Nitrogen is mainly produced by intermediate-mass ($M \sim 4-7~M_{\odot}$) AGB stars, causing nitrogen to have a slightly longer formation timescale than carbon \citep{Nomoto13}. In \textsc{Ponos}, the majority of the stellar particles are young ($\leq 200$ Myr), hence we do not expect many AGB stars. Although there may be some contribution from AGB stars, we expect cc-SNe to be the main enrichment mechanism in the simulation.

Due to the different formation timescales of carbon, nitrogen, and oxygen, assuming solar abundances would not be representative of the high-redshift Universe. In this work, we explore what effect does changing the elemental abundances in \textsc{Cloudy} have on the resulting FIR line luminosity values and ratios.
More specifically, we ran three non-solar models, which respectively assume $\mathrm{C/O}$ and $\mathrm{N/O}$ abundance ratios inferred from three different types of observational constraints: (i) JWST observations of a single, $z > 6$ galaxy; (ii) observations of local low-metallicity dwarf galaxies; and (iii) observations of metal-poor halo stars in the Milky Way. The observational results used in the three models are briefly described below, with the oxygen abundance and C/O and N/O abundance ratios summarised in Table~\ref{tab:abund}.

\subsubsection{`High-z' model using C/O and N/O measured at $z > 6$}

The launch of JWST has enabled us to spectroscopically estimate the C/O and N/O abundance ratios at $z > 6$ \citep{Arellano-Cordova22, Jones23, Isobe23, Cameron23, Topping24}.
Our first non-solar model, hereby referred to as the `High-z' model, uses the log(C/O) $= -1.01$ estimated by \cite{Jones23}, which is $\sim 5$ times lower than the solar ratio, for the $z = 6.23$ galaxy $\mathrm{GLASS\_150008}$ 
using rest-frame UV and optical OIII], CIII] and CIV emission lines detected with NIRSpec. This galaxy was observed as a part of the GLASS-JWST Early Release Science Program.
$\mathrm{GLASS\_150008}$ is smaller than \textsc{Ponos}, with a stellar mass of $M_{*} \approx 2.45\cdot 10^8 \, M_{\odot} $, and has a sub-solar oxygen abundance of $\mathrm{log(O/H) + 12 = 7.39}$. For this same source, \cite{Isobe23} measured log(N/O) $= -0.40$ using rest-frame UV NII] and NIV] lines,
with C/O and O/H values in agreement with \cite{Jones23}. This super-solar nitrogen abundance is much higher than local measurements \citep{Berg16, Berg19, Kojima17}; however, it is comparable to other recent super-solar N/O measurements at $z > 6$ \citep{Cameron23, Senchyna24, Topping24}. With the currently small numbers of C/O and N/O abundance ratios measurements at $z > 6$, $\mathrm{GLASS\_150008}$ may be an outlier and not representative of the general high-z population. We emphasise this because of the high N/O abundance ratio, given that such high nitrogen abundances would be expected in high-metallicity systems \citep{Vincenzo16}.

\subsubsection{Dwarfs model using C/O and N/O from local dwarf galaxies}

An alternative to studying high-redshift galaxies directly is to use more accessible local analogues. High-redshift galaxies are observed to be smaller and less settled, with low metallicities, which are conditions observed in low-metallicity dwarf galaxies. In this second non-solar model, which we hereby refer to as the Dwarfs model, we used the C/O and N/O abundance ratios from the combined sample of 40 metal-poor dwarf galaxies from \cite{Berg19}, with smaller masses compared to \textsc{Ponos} of $M_* = 10^6 - 10^8~M_{\odot}$. \cite{Berg19} estimated average values of $\mathrm{log(C/O) = -0.71}$ and $\mathrm{log(N/O) = -1.46}$, with an average oxygen abundance of $\mathrm{log(O/H) + 12 = 7.85}$ for this sample, using detections of the UV O$^{+2}$ and C$^{+2}$ collisionally excited lines obtained from the Hubble Space Telescope (HST) and optical N$^{+}$ lines from the Sloan Digital Sky Survey (SDSS). 

\subsubsection{Halo model using C/O from metal-poor halo stars}

Our third model, which we hereafter refer to as the Halo model, adopts the carbon and oxygen abundances from low-metallicity, high-alpha halo stars in the Milky Way. The abundances were estimated by \cite{Nissen14}, using the sample of halo stars from \cite{Nissen10} from measurements of optical spectra from the Very Large Telescope (VLT) and the Nordic Optical Telescope (NOT). High-alpha stars have a higher abundance of alpha elements compared to the iron abundance and are thought to be enriched by cc-SNe only \citep{Schuster12}. 
We selected the nine high-alpha stars with the lowest oxygen abundances, corresponding to the plateau in Fig.12 of \cite{Nissen14}, showing C/O versus O/H for their stellar samples. These halo stars have average log(O/H) $+ 12 = 8.02$ and log(C/O) = $ -0.65$, where we have converted the solar-normalised logarithmic abundances\footnote{Similarly to \cite{Nissen14}, we refer to solar-normalised logarithmic ratios as the difference between the logarithmic stellar abundance ratio and the solar abundance ratio, i.e. $\mathrm{[C/O]} \equiv \mathrm{log(C/O)_{star} - \mathrm{log(C/O)_{Sun}}} $.} in \cite{Nissen14} using their non-LTE corrected derived solar abundances in their Table 1. 
Nitrogen abundance measurements are not available for this sample, hence this model is not included in the investigation of the [NII] and [NIII] lines.

\subsection{Implementation in \textsc{Cloudy}}

As we used the \texttt{metals} command in \textsc{Cloudy} to specify the metallicity, the code assumes a default solar mixture \citep{Grevesse98, Holweger01,Allende_Prieto_2001,Allende_Prieto_2002} for its photoionisation modelling, scaling it with the scale factor given as an input value (see \textsc{Cloudy} documentation Hazy1, 7.10). 
To change abundances within the \textsc{Cloudy} models, we used the \texttt{element abundance} command, allowing us to specify the abundance of specific elements relative to hydrogen. 
Since the metallicities of the sources from \cite{Jones23}, \cite{Berg19}, and \cite{Nissen14} are lower than solar, it is necessary to scale the carbon, nitrogen, and oxygen abundances to a solar level, to ensure that the element abundances are then scaled correctly for the \textsc{Cloudy} models.
We relate the oxygen abundances from \cite{Jones23} (used in the High-z model) and from \cite{Berg19} (used in the Dwarfs model) to $Z/Z_{\odot}$ using the relation from \cite{Ma16},
\begin{equation}
    12 + \mathrm{log(O/H)} = \mathrm{log}(Z_{\mathrm{gas}}/Z_{\odot}) + 9 \label{eq:Ma},
\end{equation}
where $Z_{\odot} = 0.02$. 
Equation~\ref{eq:Ma}, which we used to scale the oxygen abundances to solar level, assumes a higher solar metallicity value than in \textsc{Cloudy}. Because of this, we get an enhancement in oxygen abundance, which will boost the [OIII] emission. We do not expect this light boost of [OIII] to have an effect on the [OIII]/[CII] and [OIII]/[NII] ratios, as the input carbon and nitrogen abundances are set using the C/O and N/O abundance ratios. The line emissions scale almost linearly with the element abundances, so the luminosity ratios should not be significantly affected, which we discuss further in Sect.~\ref{sec:effects}.

Using Eq.~\eqref{eq:Ma} results in a metallicity of log($Z_{\mathrm{gas}}$/$Z_{\odot}$) = $-1.61$ for the high-z galaxy in \cite{Jones23}, related to its measured O/H value. We get log($Z_{\mathrm{gas}}$/$Z_{\odot}$) = $-1.15$ for the dwarf galaxies in \cite{Berg19} related to their measured average oxygen abundance. Using the log($Z_{\mathrm{gas}}$/$Z_{\odot}$) ratios, we then estimate what the oxygen abundance would be at solar metallicity, which is then used as the input oxygen abundance. At solar metallicities, this results in an input oxygen abundance of $\mathrm{log(O/H) = -3.0}$ for the High-z model and the Dwarfs model.
For the Halo model, since we used measured abundances from stars and not galaxies, we used the relation between the iron abundance and the stellar metallicity from \cite{Ma16},
\begin{equation}
    [\mathrm{Fe/H}] = \log (Z_{*}/Z_{\odot}) - 0.2 \label{eq:Ma2},
\end{equation}
resulting in a metallicity of log$(Z_{*}$/$Z_{\odot})$  = $-1.035$. We used the same method of scaling the oxygen abundance, resulting in an input abundance of $\mathrm{log(O/H) = -2.95}$ at solar metallicities for the Halo model. Since Eq.~\ref{eq:Ma2} relates the iron abundance to the stellar metallicity, while Eq.~\ref{eq:Ma} relates the oxygen abundance to the gas-phase metallicity, we get a different input oxygen abundance for the Halo model when scaling the measured O/H ratio for the halo stars.
The carbon and nitrogen abundances for the High-z, Dwarfs and Halo models were then calculated using the C/O and N/O ratios of the models, based on the input oxygen abundance.
The relative abundances of the elements in the individual \textsc{Cloudy} models are constant, assuming a uniform distribution of elements, similar to the models used by \cite{Olsen17, Katz22, Pallottini22} and the fixed-temperature \textsc{Cloudy} models in \cite{Lupi2020}. We assume uniform distribution of elements since we are not tracing the evolution of the carbon, nitrogen, and oxygen abundances directly for each cell in the simulation. At $z = 6.5$, the main enrichment is due to cc-SNe, since the stellar particles are young (< $200$ Myr old). We therefore do not expect many AGB stars in the simulation, making the effect of the non-uniform heavier element distribution negligible. 
Additionally, for a low-mass galaxy ($ M_* = 2\times 10^9~M_{\odot}$) such as \textsc{Ponos}, the gas is mixed by mergers and supernovae feedback. This results in a flat metallicity gradient (slope $> -0.1~[\rm dex/kpc]$) in the star-forming galactic disc, which is in agreement with observed estimates of similar-mass galaxies at $z > 6$ \citep{Venturi24, Garcia25}.  

It is important to note that we did not change the intrinsic metallicity of \textsc{Ponos} itself, only the elemental abundances in \textsc{Cloudy}. Nevertheless, changing the elemental abundances affects the line luminosities and, consequently, the luminosity ratios.

\section{Results}\label{sec:results}

\begin{figure*}[ht]
    \centering
    \includegraphics[width=0.75\textwidth]{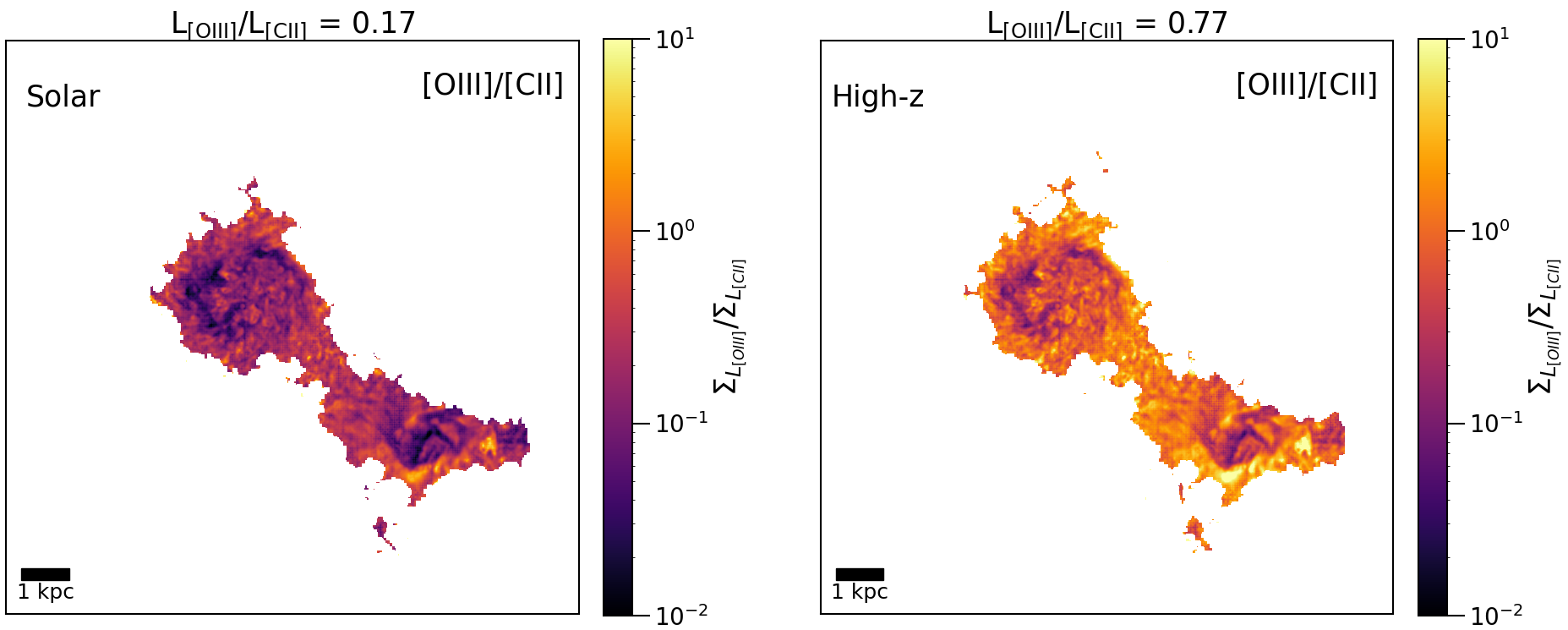}
    \caption{Surface brightness ratio maps of the $\mathrm{[OIII]/[CII]}$ luminosity ratio, comparing the Solar model (left panel) and the High-z model (right panel). The title above each panel gives the total luminosity ratios. The ratio maps obtained with the Dwarfs and Halo models are shown in Appendix~\ref{app:dwarfs}.}  
    \label{fig:ratio}
\end{figure*}

\begin{table}[]
\centering
\caption{Total [CII], [OIII], [NII], and [NIII] line luminosities.}
\footnotesize
\begin{tabular}{lccc} 
    \hline
    \hline
    Solar & Total [$L_{\odot}$] & Disc+merger B [$L_{\odot}$] & \% of tot \\
    \hline
     $\mathrm{[CII]158\mu m}$ & $4.20\cdot 10^8$ & $3.69\cdot 10^8$ & $87.79$ \\ 
     $\mathrm{[OIII]88\mu m}$  & $7.06\cdot 10^7$ & $5.40\cdot 10^7$ & $76.45$ \\
    $\mathrm{[NII]122\mu m}$ & $2.32\cdot 10^7$ & $2.00\cdot 10^7$ & $86.16$ \\ 
     $\mathrm{[NIII]57\mu m}$  & $2.48\cdot 10^7$ & $2.23\cdot 10^7$ & $90.14$ \\
    \hline
    High-z & & & \\
    \hline
     $\mathrm{[CII]158\mu m}$ & $1.81\cdot 10^8$ & $1.60\cdot 10^8$ & $88.53$ \\ 
     $\mathrm{[OIII]88\mu m}$  & $1.40\cdot 10^8$ & $1.06\cdot 10^8$ & $75.34$ \\
     $\mathrm{[NII]122\mu m}$ & $1.07\cdot 10^8$ & $9.25\cdot 10^7$ & $86.01$ \\ 
     $\mathrm{[NIII]57\mu m}$  & $1.17\cdot 10^8$ & $1.05\cdot 10^8$ & $89.99$ \\
    \hline
    Dwarfs & & & \\
    \hline
     $\mathrm{[CII]158\mu m}$ & $3.43\cdot 10^8$ & $3.02\cdot 10^8$ & $88.03$ \\ 
     $\mathrm{[OIII]88\mu m}$  & $1.42\cdot 10^8$ & $1.08\cdot 10^8$ & $75.76$\\
     $\mathrm{[NII]122\mu m}$ & $9.49\cdot 10^6$ & $8.19\cdot 10^6$ & $86.35$ \\ 
     $\mathrm{[NIII]57\mu m}$  & $1.03\cdot 10^7$ & $9.31\cdot 10^6$ & $90.16$ \\
    \hline
    Halo & & & \\
    \hline
     $\mathrm{[CII]158\mu m}$ & $4.32\cdot 10^8$& $3.79\cdot 10^8$ & $87.78$  \\ 
     $\mathrm{[OIII]88\mu m}$  & $1.59\cdot 10^8$  & $1.20\cdot 10^8$  & $75.36$ \\
     $\mathrm{[NII]122\mu m}$ & $2.33\cdot 10^7$ & $2.01\cdot 10^7$ & $86.23$ \\ 
     $\mathrm{[NIII]57\mu m}$  & $2.54\cdot 10^7$ & $2.29\cdot 10^7$ & $90.07$ \\
    \hline
\end{tabular}
\tablefoot{The first and second columns show the emission lines and the total luminosities for each of the modelling approaches. The third and fourth columns show the total luminosity from the main disc and merger B, and its percentage of the total line luminosity in the second column.}
\label{tab:lum}
\end{table}

\begin{figure}[ht]
    \centering
    \includegraphics[width=\linewidth]{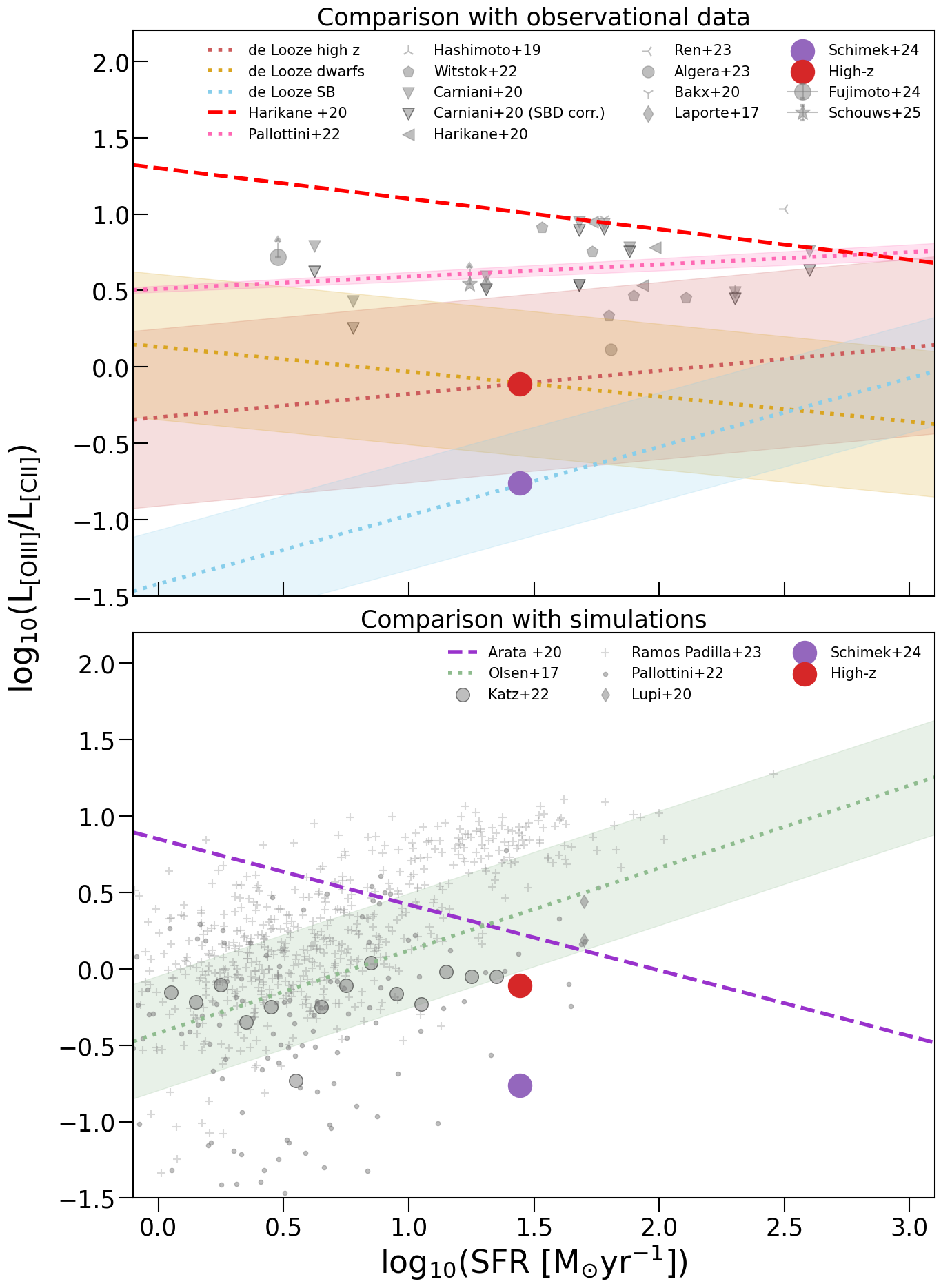}
    \caption{Total $\mathrm{[OIII]/[CII]}$ line luminosity ratio as a function of SFR for the High-z model (red) and \cite{Schimek23, Schimek24} (purple). The top panel reports the comparison with observational data from \cite{Hashimoto19} at $z \sim 7$, \cite{Witstok22} at $z = 6.7 - 7$, \cite{Carniani20} at $z \sim 6 - 9$, missing [CII] flux corrected values from \cite{Carniani20} (SBD corr.) at $z \sim 6 - 9$,  \cite{Harikane20} at $z \sim 6$, \cite{Ren23} at $z = 7.2$, \cite{Algera23} at $z = 7.3$, \cite{Bakx20} at $z = 8.31$, \cite{Laporte_2017} at $z = 8.3$, \cite{Fujimoto22} at $z = 8.5$ (lower limit), and \cite{Schouws24, Schouws25} at $z = 14.2$ (lower limit). The relations included are the $z = 6-9$ relation from \cite{Harikane20}, the best-fit relation from \cite{Pallottini22}, and the high-redshift, local dwarfs, and local starbursts (SBs) relations from \cite{deLooze14}. The bottom panel reports the comparison with simulation studies by \cite{Katz22} ($z = 6$), \cite{Padilla23} ($z = 6$), \cite{Pallottini22} ($z = 7.7$), and \cite{Lupi2020} ($z = 6$). The relations are from \cite{Arata20} ($z = 6-9$) and \cite{Olsen17} ($z \simeq 6$). }  
    \label{fig:lum_ratio_vs_SFR}
\end{figure}

\subsection{Line luminosities}
The resulting total $\mathrm{[CII]}$, $\mathrm{[OIII]}$, $\mathrm{[NII]}$, and $\mathrm{[NIII]}$ line luminosities from the \textsc{Ponos} halo, obtained by our four models (including the solar one), are listed in Table~\ref{tab:lum}. The table also lists separately the contribution from the brightest ISM components that are due to the main disc and the closest merging companion (merger B, see Fig.~\ref{fig:overview}). 

The total [CII] line luminosity in the High-z model is a factor of $2.3$ times lower compared to the Solar model, while the [OIII] luminosity is $\sim 2$ times higher than the Solar model luminosity. This results in a [OIII]/[CII] luminosity ratio of $0.77$, which is $4.5$ times higher than the Solar model ratio of $0.17$. The total [NII] and [NIII] luminosities from the High-z model are both enhanced by a factor $\sim 5$ compared to the Solar model. This results in the High-z model obtaining a [OIII]/[NII] ratio of $1.3$, which is lower than the Solar model ratio by a factor of $2.3$.

The Dwarfs and the Halo models produce total [CII] luminosity values that are not too dissimilar from the Solar model, and [OIII] luminosities that are higher by a factor of $\sim2$. This results in very similar [OIII]/[CII] luminosity ratios equal to $0.40$ and $0.37$ respectively for these two models.
The [NII] and [NIII] luminosities obtained with the Dwarfs model are both $\sim 2.4$ times lower compared to the Solar model luminosities, and the resulting [OIII]/[NII] ratio becomes very high, $\sim 15$. 

In general, our non-solar models display [CII] luminosity values that are consistent with or slightly lower than the Solar model, but they all show an increased [OIII]
luminosity, resulting in higher [OIII]/[CII] ratios compared to the Solar model. In terms of the nitrogen lines, we find large discrepancies of an order of magnitude between the total [NII] and [NIII] luminosities in the High-z and Dwarfs models, which results in vastly different [OIII]/[NII] ratios. 

The contributions from the main disc and merger B components seem to be more or less constant between all four models, which is expected since we change the element abundances according to the same abundance ratios, hence not treating the ISM and CGM components differently.

\subsection{Line emission maps}

To examine any spatial variations in the line emissions that are due to the adoption of different elemental abundances, we produced  $\mathrm{[CII]}$, $\mathrm{[OIII]}$, $\mathrm{[NII]}$ and [NIII] line maps with the four models. Since both the [NII] and [NIII] lines trace ionised gas, we only show the [NII] maps here; however, the analysis of [NII] is applicable to [NIII].

In Fig.~\ref{fig:CII_maps} we compare the $\mathrm{[CII]}$ line emission maps obtained with the Solar model (top panels) and High-z model (bottom panels).
The weaker [CII] emission seen in the High-z model map reflects its overall lower [CII] line luminosity compared to the Solar model. The High-z model [CII] map shows a lower surface brightness in the central ISM components of the disc and merger B, as well as lower extended emission in the tidal tails (and so, lower emission from CGM components). The Dwarfs and Halo model [CII] maps are shown in Appendix~\ref{app:dwarfs}: the former shows a slightly weaker emission compared to the Solar model, while the latter is more similar to the Solar model map.

In Fig.~\ref{fig:OIII_maps} we show the [OIII] line emission maps for the Solar model (top panels) and the High-z model (bottom panels). The High-z model shows a higher [OIII] surface brightness in the central ISM regions and stronger extended emission compared to the Solar model. This reflects the higher total [OIII] luminosity in the High-z model. We find similar results for the Dwarfs and Halo models [OIII] maps in Appendix~\ref{app:dwarfs}.

In Fig.~\ref{fig:NII_maps} we compare the [NII] line emission maps obtained with the Solar (top panels), High-z (middle panels), and Dwarfs (bottom panels) models. 
The [NII] surface brightness in the High-z model map is significantly higher in the ISM compared to the Solar model map, in addition to exhibiting stronger extended emission. The Dwarfs model map shows the opposite trend, with a lower central [NII] surface brightness and much weaker extended emission compared to the Solar model. The large discrepancy in total [NII] luminosity between the High-z and Dwarfs models is reflected in their vastly different [NII] maps.

\subsection{[OIII]/[CII] ratio maps} \label{sec:ratio_maps}
 
Before computing the [OIII]/[CII] ratio maps, we masked out the weak emission for both [OIII] and [CII] by setting a minimum line surface brightness threshold of $\Sigma_{\mathrm{line}} < 1$ $L_{\odot}$ kpc$^{-2}$, which is low and has a negligible effect on the total luminosity ratio value,
as discussed in \cite{Schimek24}.
This allows us to focus our study of the [OIII]/[CII] ratio on the brightest ISM components of \textsc{Ponos}, that is the main disc and merger B, which account for $\gtrsim 80\%$ of the total line luminosities (see Table~\ref{tab:lum}). 

In Fig.~\ref{fig:ratio} we show the [OIII]/[CII] line ratio maps obtained from the Solar model (left panel) and the High-z model (right panel).
The factor $4.5$ higher [OIII]/[CII] luminosity ratio obtained by the High-z model is clearly reflected in the ratio maps.
The distribution of the high-ratio and low-ratio regions in both the Solar model map and the High-z model map occur at the same locations. This is expected from the modelling, as we change the carbon and oxygen abundances according to the same C/O ratio for all individual \textsc{Cloudy} models in each grid. 
The Dwarfs model and Halo model ratio maps in Appendix~\ref{app:dwarfs} reflect their higher luminosity ratios compared to the Solar model, in addition to showing the same spatial distributions as the Solar model map. The luminosity ratio in the High-z, Dwarfs and Halo models all seem to change uniformly throughout the maps. 

\section{Discussion}\label{sec:discussion}

In this section, we discuss our results and compare them with 
previous observational and theoretical works.

\subsection{The [OIII]/[CII] luminosity ratio}

In Fig.~\ref{fig:lum_ratio_vs_SFR} we show the [OIII]/[CII] luminosity ratio as a function of SFR obtained from our High-z model (red), together with the value obtained by \cite{Schimek23} by adopting solar abundances (purple). 
The top panel compares our results to observations of $z > 6$ galaxies and local analogues, while the bottom panel shows our results compared to other theoretical efforts. We show only the High-z model out of our three non-solar models because it is the one producing the highest [OIII]/[CII] values.

Focusing on the top panel, our High-z model delivers a [OIII]/[CII] ratio that is a factor of $4.5$ higher than the ratio from \cite{Schimek23}, and so it is closer to the observed $z > 6$ data. Furthermore, the High-z model is in good agreement with the high-redshift ($z = 0.5-6.6$) and local ($z < 0.5$) metal-poor dwarf relations from \cite{deLooze14}. However, despite the improvement, the High-z model is still falling short of most of the [OIII]/[CII] luminosity ratios observed at $z > 6$, most notably it is a factor of 13 lower than the $z = 6-9$ estimates from \cite{Harikane20}. The discrepancy with the observations is reduced significantly if comparing the High-z model ratio to the \cite{Pallottini22} relation: a best-fit relation of the $z = 6-9$ observational data from \cite{Carniani20} that corrects for missing [CII] flux, i.e. surface brightness dimming (SBD), at low SFR.
For what concerns the individual [CII] and [OIII] line luminosities, the High-z model is in very good agreement with observational data on the [CII]-SFR relation, while it sits below most of the data points on the observed [OIII]-SFR relation (see plots reported in the Appendix~\ref{app:SFR}).
We note that most of the available observational constraints are for sources at higher SFRs than \textsc{Ponos}. As there is only a small sample of [OIII] observations compared to [CII], it is possible that the observed [OIII]/[CII] luminosity ratios are biased towards the brightest [OIII]-emitters \citep{Bakx24}. 

Focusing on the bottom panel, we compute a lower ratio than \cite{Padilla23} and are barely outside the scatter of \cite{Olsen17}, which is likely due to them producing significantly lower [CII] emission at the comparable SFR (see Fig.~\ref{fig:CII_vs_SFR}).
The High-z model is close to the $z = 6-9$ simulated estimates from \cite{Arata20}, who track the carbon and oxygen abundances for each particle and include enrichment from cc-SNe, Type Ia SNe and AGB stars. This way, the enrichment timescales for carbon and oxygen are included more accurately. 
Our modelling is in very good agreement with the $z = 6$ data from \cite{Katz22} (with a spatial resolution of $\mathrm{7.3\, pc\, h^{-1}}$ similar to the \textsc{Ponos} simulation), at comparable SFR. The agreement in [OIII]/[CII] ratio values is expected, as we in this work apply a low C/O abundance ratio from observational constraints consistent with assumptions of pure cc-SNe enrichment \citep{Jones23}. Comparing the [CII]-SFR and [OIII]-SFR relations at similar SFR as \textsc{Ponos}, \cite{Katz22} line luminosities are lower than what we obtain from our High-z model (see Fig.~\ref{fig:CII_vs_SFR} and Fig.~\ref{fig:OIII_vs_SFR}).

\subsection{The C/O abundance ratio and metallicity as contributing factors
to the [OIII]/[CII] ratio}

\begin{figure}[ht]
    \centering
    \includegraphics[width=\linewidth]{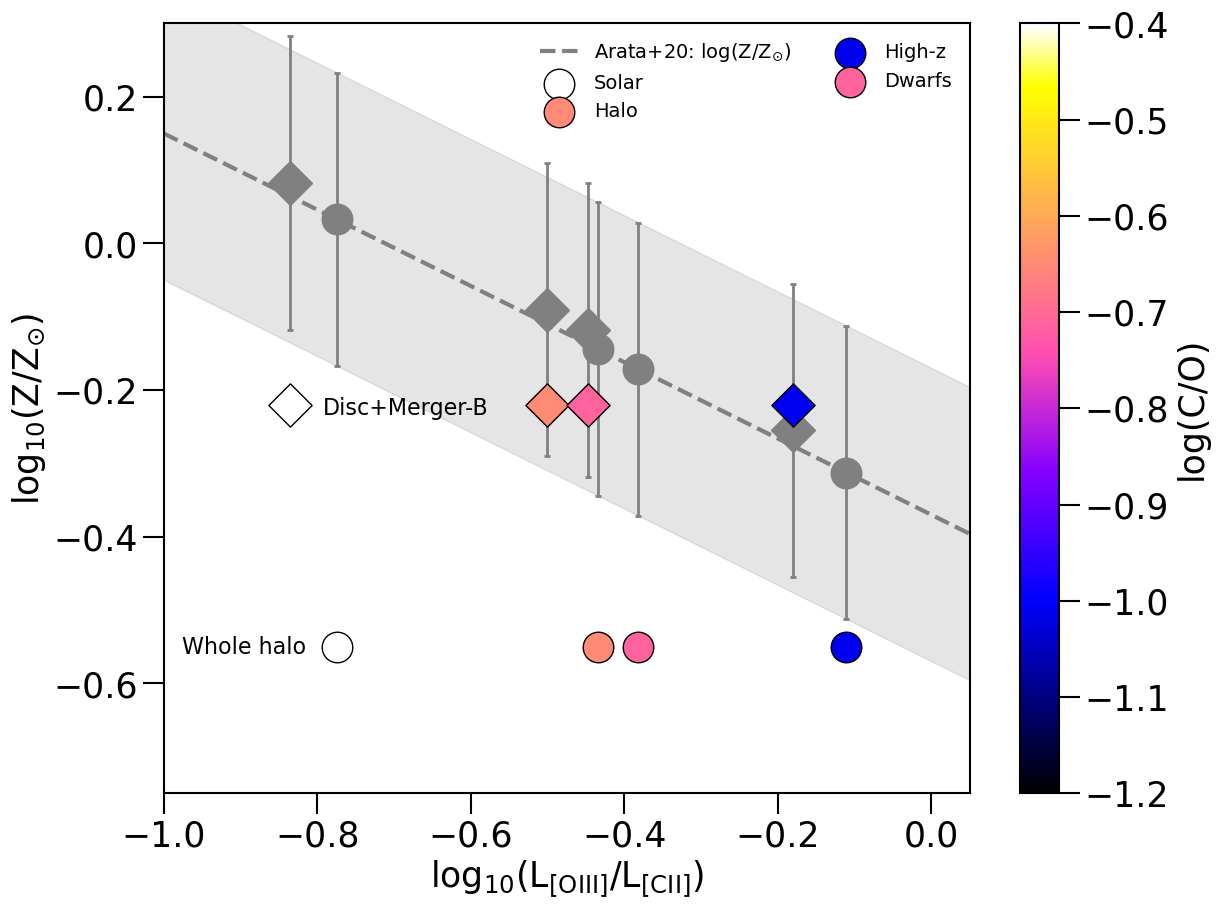}
    \caption{$\mathrm{[OIII]/[CII]}$ luminosity ratio vs metallicity $Z/Z_{\odot}$ in \textsc{Ponos}. Our models are colour-coded based on the assumed $\mathrm{C/O}$ abundance ratios. The circle symbols correspond to the luminosity ratios and average metallicities computed for the whole simulated halo ($\log(Z/Z_{\odot}) = -0.55$). The diamonds show the contribution from the main disc and merger B component only ($\log(Z/Z_{\odot}) = -0.22$). 
    The grey dashed line is the best-fit relation from \cite{Arata20} for their simulated sample of $z = 6-9$ galaxies, with the grey area corresponding to their $\lesssim 0.2$ dex dispersion. The grey symbols are the \textsc{Ponos} models shifted to the metallicity values expected from the \cite{Arata20} relation.}
    \label{fig:metallicity}
\end{figure}

With the lowest C/O abundance ratio of all our models, the High-z model obtained the highest [OIII]/[CII] ratio. The trend of increasing [OIII]/[CII] ratio for lower C/O abundance values is consistent with the theoretical predictions from \cite{Katz22} and \cite{Arata20}. \cite{Arata20} also found that the C/O ratio decreases at higher redshifts. \cite{Harikane20} found, however, using \textsc{Cloudy} modelling, that lower C/O abundances can only partially explain the highest [OIII]/[CII] luminosity ratios observed at $z > 6$. 

To further explore the significance of including a physically motivated C/O abundance ratio, we plot in Figure~\ref{fig:metallicity} the [OIII]/[CII] luminosity ratios obtained from our models as a function of the average intrinsic metallicity of \textsc{Ponos}, and compare them with the [OIII]/[CII]-metallicity relation from \cite{Arata20} (grey dashed line),
\begin{equation}\label{eq:arata}
    \log(Z/Z_{\odot}) = -0.37 - 0.52\log([\rm OIII]/[\rm CII]). 
\end{equation}
Equation~\eqref{eq:arata} is a best-fit relation derived by \cite{Arata20} using their simulated $z = 6-9$ sample with luminosity ratios $-0.5 < \mathrm{log([OIII]/[CII])} < 2.0$.
\cite{Arata20} were motivated by the need for estimating the observed gas-phase metallicities at high redshifts without using iron tracers that do not consider the enhancement of $\alpha$-elements in early galaxy evolution.
As explained above, our methodology consists of changing the element abundances in \textsc{Cloudy}, and so all models share the same intrinsic metallicity of the \textsc{Ponos} simulation, equal to log$(Z/Z_{\odot}) = -0.55$ (circles in Fig.~\ref{fig:metallicity}), and an average metallicity of log$(Z/Z_{\odot}) = -0.22$ when considering only the main disc and merger B, i.e. removing the more extended CGM components and the merger A (diamonds in Fig.~\ref{fig:metallicity}). 
The colour coding in Fig.\ref{fig:metallicity} corresponds to the C/O abundance ratios in our models and the grey symbols represent the metallicities of our models extrapolated from the [OIII]/[CII] luminosity ratios by assuming the relation by \cite{Arata20}.
Compared to the other models, we find that the High-z model is in better agreement with the expectations from the \cite{Arata20} relation, and it is perfectly consistent with this relation when considering only the main disc and merger B. Therefore, we can conclude that using a C/O abundance ratio from observations of a $z > 6$ galaxy produces [OIII]/[CII] luminosity ratios that are more consistent with the intrinsic metallicity of \textsc{Ponos}. 

\subsection{The ionisation parameter as a contributing factor to the [OIII]/[CII] ratio}

The High-z model estimates a lower [OIII] line luminosity than high-redshift observations and theoretical modelling efforts from \cite{Arata20} at comparable SFR (see e.g. Fig.~\ref{fig:OIII_vs_SFR}), which may contribute to the fact that the  [OIII]/[CII] luminosity ratio is still lower than most observations at $z > 6$. For their simulations, \cite{Arata20} estimate high ionisation parameters of $-2.2 \lesssim \mathrm{log}(U) \lesssim -1.5$, consistent with recent high-redshifts observations \citep{Bunker23,Topping24, Fujimoto22}. Higher ionisation parameters have also been discussed as contributing factors to the enhanced [OIII]/[CII] luminosity ratio in observational studies \citep{Harikane20}. For comparison with these works, we estimate the log($U$) value of \textsc{Ponos} using the volume-average estimate from \cite{Hirschmann17}:
\begin{equation}
   U = \frac{3\alpha_{\rm{B}}^{2/3}}{4c} \left( \frac{3Q_{\rm{ion}}\epsilon^2  n_{\rm{H},*}}{4\pi}    \right)^{1/3} \label{eq:U},
\end{equation}
where $Q_{\rm{ion}}$ is the rate of ionising photons which we take from the RT post-processed simulation, $n_{\rm{H},*} = 10^2\, \rm{cm}^{-3}$ is the hydrogen number density in HII-regions used in \cite{Hirschmann17}, $\epsilon = n_{\rm{gas}}/n_{\rm{H},*}$ is the filling factor, and $\mathrm{\alpha_B}$ is the case-B hydrogen recombination coefficient.
For the whole galaxy halo, we estimate log($U$) = $-2.6$, and for the main disc and merger B components, we estimate log($U$) $= -2.45$. These are lower than what was estimated by \cite{Arata20} for their sample, hence consistent with the hypothesis that a higher ionisation parameter would result in a higher [OIII] luminosity. 
Using \textsc{Cloudy} models, \cite{Harikane20} found high ionisation parameters of log($U$) $=-1$ to reproduce their highest [OIII]/[CII] ratios, which is significantly higher than our log($U$) estimate for \textsc{Ponos}, even when we consider only the main disc and merger B. 
Similarly high parameters have been found in observations by \cite{Topping24} for a $z \approx 6$ galaxy. From observations of GN-z11 at $z = 10.6$, \cite{Bunker23} computed log($U$) $=-2.25$, which is only a factor of $\sim2$ higher than \textsc{Ponos}. 

We thus conclude that higher ionisation parameters resulting from extreme high-redshift environments, for example top-heavy initial mass functions, or spectral energy distributions with harder radiation fields, could contribute to enhancing the [OIII]/[CII] luminosity ratio further.

\subsection{The nitrogen lines and the [OIII]/[NII] ratio}

\begin{figure}[ht!]
    \centering
    \centerline{\includegraphics[width=\linewidth]{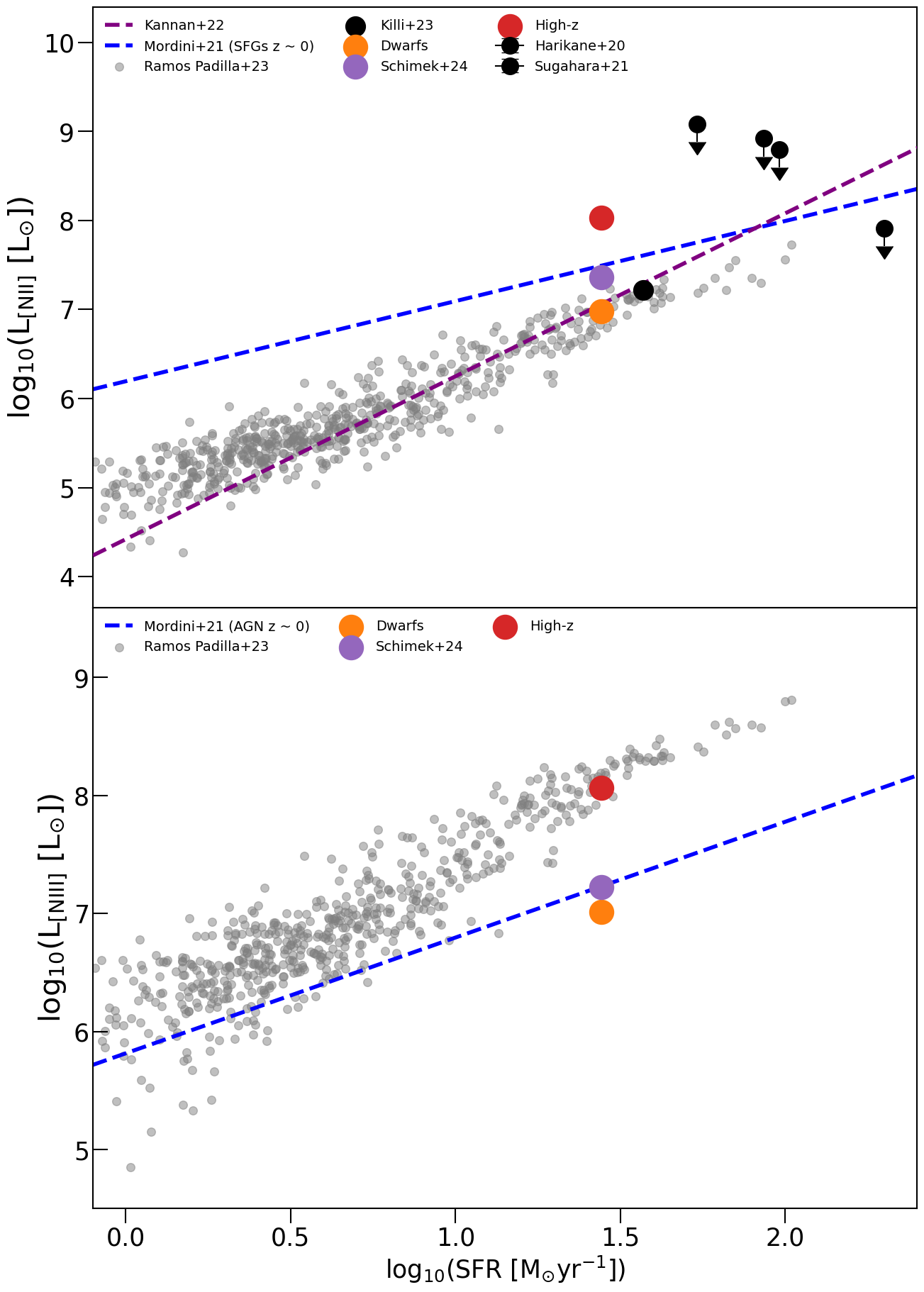}}
    \caption{Total $\mathrm{[NII]}$ luminosity (top panel) and $\mathrm{[NIII]}$ luminosity (bottom panel) as functions of SFR for the High-z model (red), Dwarfs model (orange), and \cite{Schimek23, Schimek24} (purple). 
    Both panels include comparisons with the simulated $z = 6$ sample from \cite{Padilla23} (grey circles) and the linear relations from \cite{Mordini21} based on observations of galaxies in the local Universe. In the top panel we include the [NII] observation from \cite{Killi23} (black circle) at $z \sim 7$, the observed upper limits from \cite{Harikane20} at $z \sim 6$ and \cite{Sugahara21} at $z \sim 7$ (black circles with arrows), and the theoretical relation from \cite{Kannan22} ($z = 6-10$). 
    }  
    \label{fig:NII_NIII_vs_SFR}
\end{figure}

\begin{figure}[ht]
    \centering
    \centerline{\includegraphics[width=\linewidth]{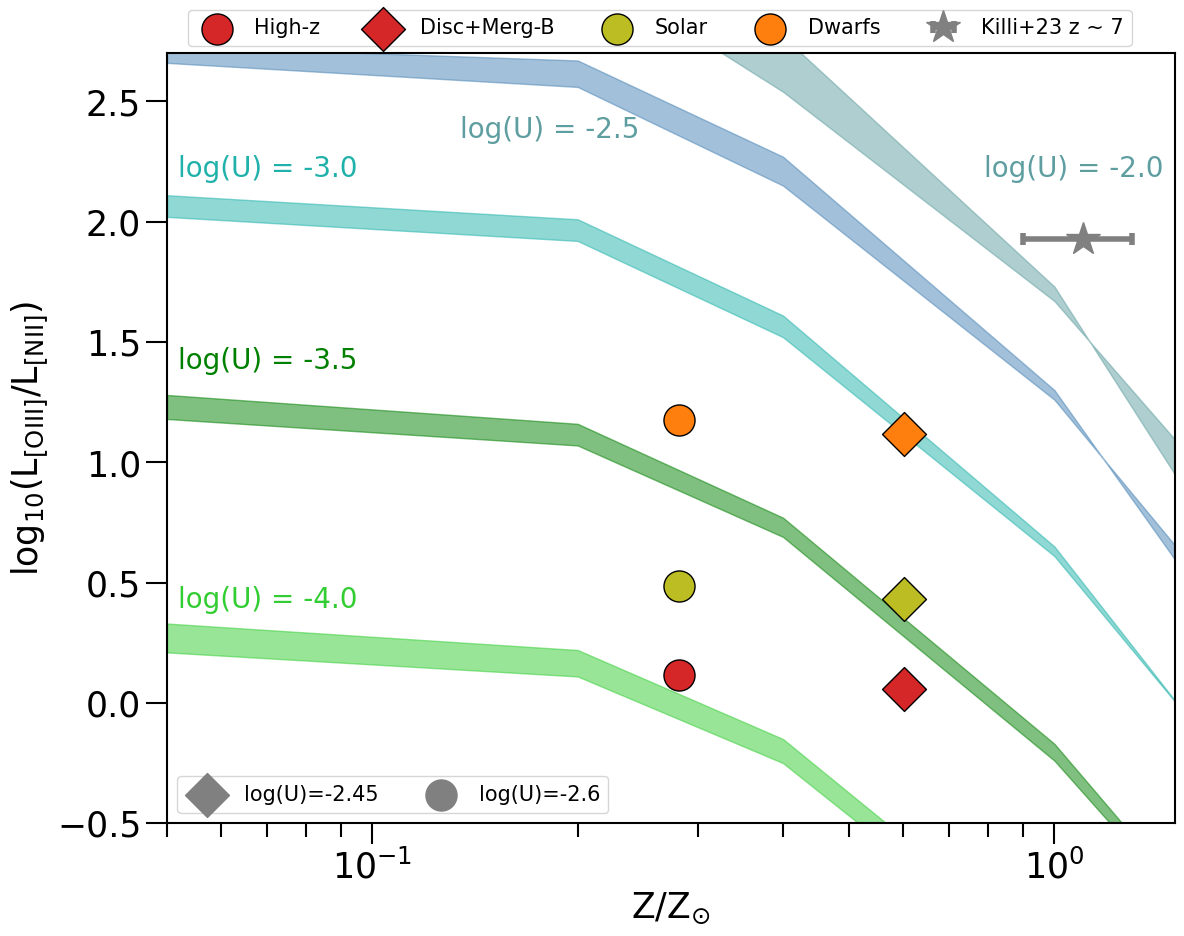}}
    \caption{[OIII]/[NII] as a function of metallicity $Z\, [Z_{\odot}]$ for the Solar model (yellow-green), High-z model (red) and Dwarfs model (orange). The circles show the contribution from the whole galactic halo, while the diamonds consider only the main disc and merger B. The shaded regions are the model predictions of the [OIII]/[NII] ratio as functions of metallicity from \cite{Pereira17}, with each shaded region corresponding to a model assuming a certain ionisation parameter: $\log(U) = -4, -3.5, -3, -2.5$ and $\mathrm{-2}$. The width of the shaded regions spans the density range: $\log(n_{\rm{H}}) = 1-4$. The grey star represents the observed ratio from \cite{Killi23} at $z \sim 7$.}  
    \label{fig:OIII_NII_Z}
\end{figure}

Observations of [NII] and [NIII] at $z > 6$ are very limited (see Figure~\ref{fig:NII_NIII_vs_SFR}), resulting in a lack of observational constraints adequate to test the results of our models. Nevertheless, it is interesting to explore further the [OIII]/[NII] ratio because of its potential as a metallicity tracer at high-z \citep{Nagao11, Pereira17, Rigo18}.

In the top panel of Fig.~\ref{fig:NII_NIII_vs_SFR} we plot the total [NII] line luminosity as a function of SFR obtained by our High-z (red) and Dwarfs (orange) models and \cite{Schimek23, Schimek24} (purple). For comparison with observations, we include the $z \sim 7$ observation from \cite{Killi23} (black circle), which is the first detection of [NII] in a non-quasar galaxy at $z > 6$, and the observed upper limits at $z > 6$ from \cite{Harikane20} and \cite{Sugahara21} (black circles with arrows). Due to the lack of high-redshift observations, we have also included the relation from \cite{Mordini21} based on observations of local star-forming galaxies (SFG), where we adopted the \cite{Kennicut_Evans12} conversion between $L_{\rm{IR}}$ and SFR similar to \cite{Padilla23}. In the same panel, we include the $z = 6$ simulations from \cite{Padilla23} (grey circles) and the theoretical relation from \cite{Kannan22}. Both \cite{Padilla23} and \cite{Kannan22} have resolutions at $m_{\rm{gas}} \geq 10^5~M_{\odot}$, significantly lower than \textsc{Ponos}. 
The total [NII] luminosity resulting from the High-z model, for which we adopted $\mathrm{log(N/O) = -0.40}$ \citep{Isobe23}, is higher than the $z \sim 7$ observational constrain from \cite{Killi23} and the relation from \cite{Mordini21} at comparable SFR to \textsc{Ponos}. The total [NII] luminosity is also higher than the theoretical predictions from \cite{Padilla23} and \cite{Kannan22} at comparable SFR. We stress that the super-solar N/O abundance ratio adopted in the High-z model is based on recent JWST observations of a single $z > 6$ galaxy \citep{Isobe23}, which might not be representative of the broader population of high-redshift galaxies. Recent $z > 6$ data from JWST have uncovered a few surprisingly high N/O abundance ratios \citep{Cameron23, Senchyna24, Topping24}; however, the origin of these high abundance ratios is still unknown.
Our predictions from the Dwarfs model, with sub-solar $\mathrm{log(N/O) = -1.46}$ \citep{Berg19}, and the prediction from \cite{Schimek23}, are in better agreement with the observational data from \cite{Killi23} and \cite{Mordini21}. The Dwarfs model and the fiducial Solar model by \cite{Schimek23} are also consistent with the theoretical estimates from \cite{Padilla23} and \cite{Kannan22}.

In the bottom panel of Fig.~\ref{fig:NII_NIII_vs_SFR} we plot the total [NIII] line luminosities as functions of SFR for our High-z model (red), Dwarfs model (orange) and \cite{Schimek23} (purple). We include a relation from \cite{Mordini21} based on local Active Galactic Nucleus (AGN) observations, again adopting the \cite{Kennicut_Evans12} conversion. We compare to AGN data, as there are no available non-AGN emission line measurements.
For theoretical comparison, we have included the $z = 6$ simulations from \cite{Padilla23} (grey circles). 
We find that the total [NIII] luminosity obtained by our High-z model is in very good agreement with the \cite{Padilla23} simulations at comparable SFR, while the lower values from the Dwarfs model and from \cite{Schimek23} agree with the local relation from \cite{Mordini21}. We remind that, the total [NIII] luminosity in our Solar model is $\sim 50 \%$ higher than in \cite{Schimek23} due to the addition of high-temperature (log$(T) = 5.7, 6.0$) grid values in \textsc{Cloudy} (see Sect.~\ref{sec:solar_coarser}). Therefore, our [NIII] predictions should be considered as upper limits. 

Comparing our results from the top and bottom panels of Fig. ~\ref{fig:NII_NIII_vs_SFR}, we find that none of our models agrees with both the [NII]-SFR and [NIII]-SFR relations predicted by \cite{Padilla23}. 
Due to the small sample of nitrogen line observations available at high-z, the question regarding the behaviour of the [NIII]-SFR relation in the early Universe remains open.

Since $\mathrm{O^{++}}$ and $\mathrm{N^+}$ have critical densities of $\mathrm{510~cm^{-3}}$ and $\mathrm{310~cm^{-3}}$ respectively, the [OIII]/[NII] ratio has a low dependence on the ISM gas density. However, this luminosity ratio is highly dependent on the ionisation parameter. 
In Fig.~\ref{fig:OIII_NII_Z}, we show the [OIII]/[NII] ratio as a function of the intrinsic metallicity of \textsc{Ponos} for the solar (yellow-green), High-z (red) and Dwarfs (orange) models. The circles show the contribution from the whole galaxy halo, while the diamonds consider only the brightest components (main disc and merger B). For comparison, we have included the theoretical predictions from \cite{Pereira17}, where each shaded region corresponds to a model assuming a specific ionisation parameter: log$(U) = -4, -3.5, -3, -2.5$ and $\mathrm{-2}$. The width of the shaded regions spans the density range: log $n_{\rm{H}}$ (cm$^{-3}$) = $1-4$, showing the low dependence on gas density. For comparison to observations, we include the $z \sim 7$ observation from \cite{Killi23}.
We find that the large discrepancy in [NII] luminosity between the High-z model and the Dwarfs model, results in vastly different [OIII]/[NII] ratios. This is the result of the different N/O abundance ratios assumed.

From Fig.~\ref{fig:OIII_NII_Z}, it is clear that our models produce [OIII]/[NII] luminosity ratios well below what would be expected from the \cite{Pereira17} predictions considering the ionisation parameter of \textsc{Ponos}. 
The very low luminosity ratio obtained by the High-z model is due to the observed super-solar N/O ratio \citep{Isobe23} assumed in our modelling. We emphasise again that recent observations of $z > 6$ galaxies have found excess nitrogen that is not trivial to explain \citep{Cameron23,Isobe23, Topping24}, and we need more observational data of the [NII] emission line at high redshifts to further constrain the [OIII]/[NII] ratios.

\subsection{Other effects on the far-infrared luminosity ratios} \label{sec:effects}

The [OIII]/[CII] luminosity ratio obtained by the High-z model is $4.5$ times higher than the ratio obtained by the Solar model, while the C/O abundance ratio used as input for the High-z model is $5.1$ times lower than the input C/O in the Solar model. While the difference between the ratios is small ($\sim 10\%$), it shows that the line luminosities scale close to linear with the element abundance. However, there is something else affecting the line luminosities. 
Similar discrepancies between emission line ratios and abundance ratios were found by \cite{McClymont24}, who modelled optical emission lines arising from the diffuse ionised gas (DIG) in a simulated Milky Way-like galaxy. \cite{McClymont24} found that the relative abundance ratios were not enough to explain the behaviour of the optical line ratios. 

We investigated a few selected \textsc{Cloudy} models from our four models separately, focusing on how the relative line emissions as functions of depth behave compared to what we expect from the relative element abundances. Focusing on the [OIII] line, we found that for low-density ($n = 1$ cm$^{-3}$), high-temperature ($T = 10^{4}$ K) \textsc{Cloudy} models, the relative emission obtained by the High-z, Dwarfs and Halo models with respect to the Solar model, is slightly lower than expected, especially at depths corresponding to length of the smallest AMR grid cells. At the same temperature, but higher densities ($n = 100$ cm$^{-3}$), we find the relative [CII] emission to be higher than expected, especially for the Dwarfs and High-z models. These discrepancies increase with increasing gas-phase metallicities $Z$, as more metals are present in the gas. Furthermore, the depths at which the discrepancies are the highest decrease with increasing metallicity, which could be related to the PDR column density being proportional to $1/Z$ \citep{Kaufman06, Harikane20, Katz22}, and thus, the interface between the HII regions and PDRs is shifted. Since the line luminosities are computed considering the AMR structure in \textsc{kramses-rt} (Sect. \ref{sec:cloudy-pp}), the non-uniform scaling of the emission along the depth of the slabs affect the total line luminosities.

Note that we have only investigated a small sample of different parameters in our \textsc{Cloudy} models, and the discrepancy between the [OIII]/[CII] luminosity ratio and the C/O abundance ratio is small. While analysis of other parameters would be necessary to further investigate the discrepancy amongst the FIR lines, it is beyond the scope of this work. 

\section{Conclusions}\label{sec:conclusion}

In this work, we investigated the reasons behind the high $\mathrm{[OIII]/[CII]}$ luminosity ratios measured in high-redshift galaxies, which are significantly higher (ratios of 1-20) compared to the local Universe (ratios of $\leq$1). Additionally, we modelled the FIR nitrogen lines and studied the $\mathrm{[OIII]/[NII]}$ ratio, which is a potential metallicity tracer. We modelled FIR emission lines with \textsc{Cloudy} for the high-resolution \textsc{Ponos} simulation \citep{Ponos, Schimek23, Schimek24}, changing elemental abundance ratios to be more representative of the chemical composition found in high-redshift galaxies. We explored different input C/O and N/O abundance values, basing our assumptions on (i) JWST observations of a single high-z galaxy \citep{Jones23, Isobe23}, (ii) local low-metallicity dwarf galaxy analogues \citep{Berg19}, and (iii) observations of metal-poor halo stars in the Milky Way \citep{Nissen14}. 
Our main results can be summarised as follows:

 \begin{itemize}
      \item Reducing the $\mathrm{\log(C/O)}$ abundance ratio by a factor of 5.1 (High-z model) increases the $\mathrm{[OIII]/[CII]}$ luminosity ratio by a factor of 4.5 when compared to assuming solar abundances. 
      Additionally, our obtained total [CII] and [OIII] luminosity values are consistent with and closer to observations and simulations with similar SFR as \textsc{Ponos}.
      This shows why it is important to consider the chemistry of high-redshift galaxies when modelling their emission lines. 
      \item The [OIII]/[CII] luminosity ratio we obtain when assuming a high-redshift $\mathrm{\log(C/O)}$ abundance ratio, is closer to the luminosity ratio that we would expect based on the intrinsic metallicity of \textsc{Ponos}, assuming the theoretical [OIII]/[CII]-$Z$ relation by \cite{Arata20}.
      \item Our models produce $\mathrm{[OIII]/[NII]}$ ratios that are lower than what we expect from current theoretical models, considering the ionisation parameter we calculated for \textsc{Ponos}
      \item The large discrepancy in the N/O abundance ratios used as input in our models, from observations of a $z > 6$ galaxy and local dwarf galaxies, resulted in vastly different $\mathrm{[OIII]/[NII]}$ luminosity ratios and total [NII] and [NIII] luminosity values. Nevertheless, we obtain total luminosity values that agree with the few observational and theoretical constraints at similar SFR as \textsc{Ponos}. 
      \item Small non-linearities between the total line luminosities and the element abundances could hint towards a complex interplay of the chemistry and other ISM properties.
\end{itemize}
Our results highlight the importance of further improving and developing theoretical models of FIR and sub-mm line emission from high-z galaxies. These modelling efforts need to be coupled with high-resolution zoom-in cosmological simulations that resolve the physical processes and the small scales that are relevant to the cold and warm phase traced by FIR and sub-mm lines, not only on the ISM but also in the CGM, where a significant fraction of the [CII] emission can arise (see also \citealt{Schimek23}). These efforts are urgently needed to inform ongoing ALMA follow-ups of the increasing number of $z>6$ galaxies that are being discovered and, in the future, to interpret the wide-field, high-sensitivity line observations and surveys that will be performed by AtLAST (see \cite{Mroczkowski+24} for the telescope design, and \cite{vanKampen+24} and \cite{Lee+24} for the science drivers relevant to this study).

\begin{acknowledgements}
    This project has received funding from the European Union's Horizon Europe and Horizon2020 research and innovation programmes under grant agreements No. 101188037 (AtLAST2) and No. 951815 (AtLAST). Views and opinions expressed are however those of the author(s) only and do not necessarily reflect those of the European Union or European Research Executive Agency. Neither the European Union nor the European Research Executive Agency can be held responsible for them. C.T acknowledges support from the Knut and Alice Wallenberg Foundation.
    The simulations were performed using the resources from the National Infrastructure for High Performance Computing and Data Storage in Norway, UNINETT Sigma2, allocated to Project NN9477K.
      We acknowledge the use of the programming language Python \citep{van1995python,python3} and packages: NumPy \citep{Walt2011}, Matplotlib \citep{Hunter07} and Pynbody \citep{Pynbody}.
\end{acknowledgements}

\bibliographystyle{aa}
\bibliography{ref}

\begin{appendix}

\section{The [CII]-SFR and [OIII]-SFR relations}\label{app:SFR}
\begin{figure}[ht!]
    \centering
    \includegraphics[width=0.98\linewidth]{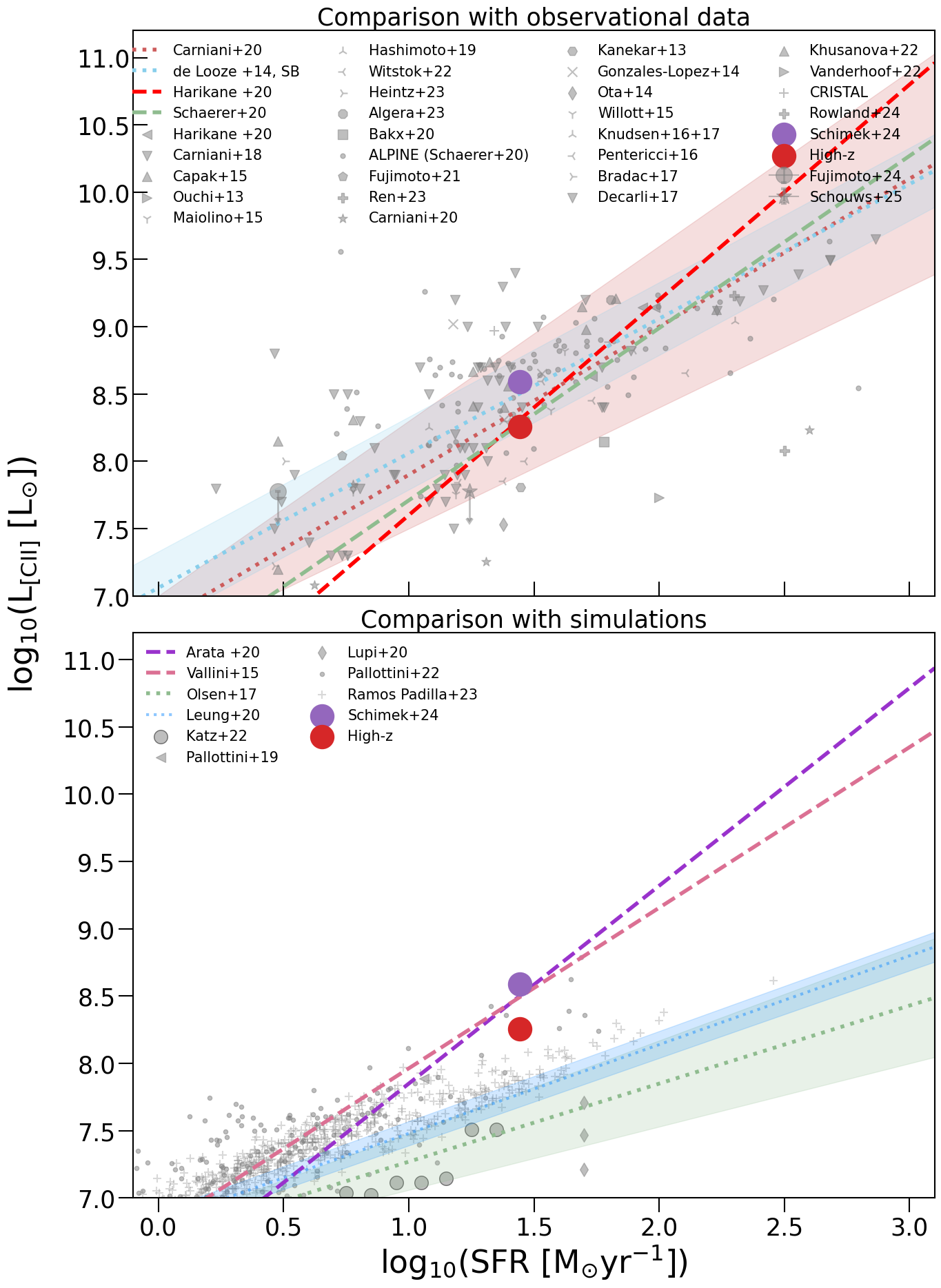}
    \caption{Total $\mathrm{[CII]}$ luminosity as a function of SFR for the High-z model (red) and \cite{Schimek23, Schimek24} (purple). In the top panel we compare our models to observational values from \cite{Harikane20} at $z \sim 6$, \cite{Carniani18} at $z = 5-7$, \cite{Capak15} at $z = 5.1- 5.7$, \cite{Ouchi13} at $z \simeq 6.6$, \cite{Maiolino15} at $z = 6.8 - 7.1$, \cite{Hashimoto19} at $z = 7.15$, \cite{Witstok22} at $z = 6.7 - 7$, \cite{Heintz23} at $z = 8.5$, \cite{Algera23} at $z = 7.3$, \cite{Bakx20} at $z = 8.31$, \cite{Schaerer20} (ALPINE) at $z = 4-8$, \cite{Fujimoto21} at $z = 6$, \cite{Ren23} at $z = 7.2$, \cite{Carniani20} at $z \sim 7-9$, \cite{Kanekar13} at $z = 6.56$, \cite{González-López_2014} at $z = 6.5-11$, \cite{Ota14} at $z = 6.9$, \cite{Willott15} at $z = 6$, \cite{Knudsen16, Knudsen17} at $z = 6-7.6$, \cite{Pentericci16} at $z = 6.6-7.1$, \cite{Bradac17} at $z = 6.7$, \cite{Decarli17} at $6-6.6$, \cite{Khusanova22} at $z = 6$, \cite{Vanderhoof22} at $z = 4.5$, CRISTAL \citep{CRISTAL_Posses24} at $z = 5.5$, \cite{Rowland24} at $z = 7.31$ , \cite{Fujimoto22} at $z = 8.5$ (upper limit), and \cite{Schouws24,Schouws25} at $z = 14.2$ (upper limit). We include relations from \cite{Carniani20} at $z > 6 $, \cite{deLooze14} for local SB galaxies, \cite{Harikane20} at $z = 6-9$ and \cite{Schaerer20} at $z = 4-8$. In the bottom panel, we compare our models to simulated values from \cite{Katz22} ($z = 6$), \cite{Pallottini19} ($z = 6$), \cite{Lupi2020} ($z = 6$), \cite{Pallottini22} ($z = 7.7$), and \cite{Padilla23} ($z = 6$). The relations are from \cite{Arata20} ($z = 6-9$), \cite{Vallini15} ($z \approx 7 $), \cite{Olsen17} ($z \simeq 6$), and \cite{Leung20} ($z \simeq 6$).} 
    \label{fig:CII_vs_SFR}
\end{figure}

\begin{figure}[ht!]
    \centering
    \includegraphics[width=0.98\linewidth]{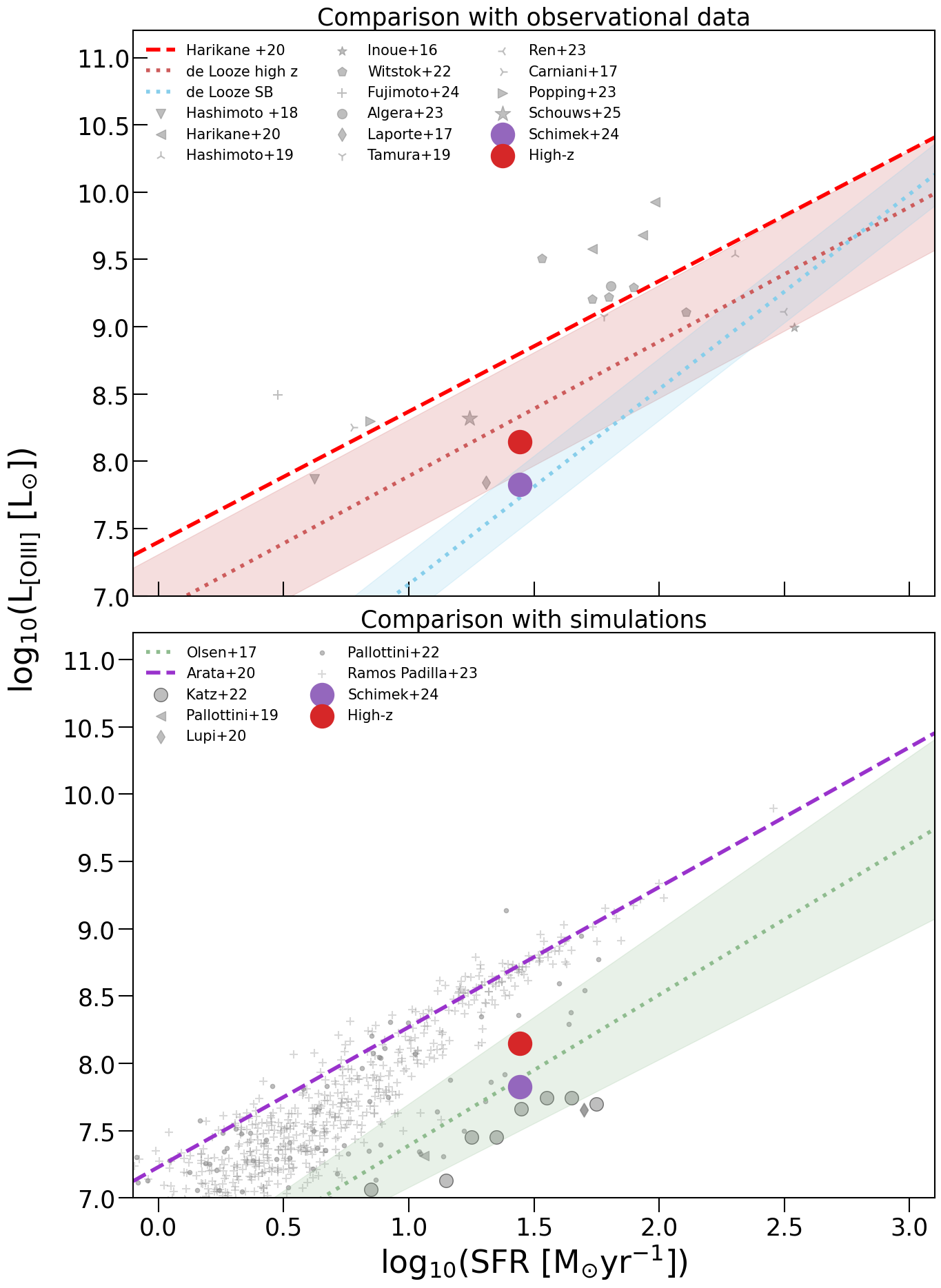}
    \caption{Total $\mathrm{[OIII]}$ luminosity as a function of SFR for the High-z model (red) and \cite{Schimek23, Schimek24} (purple). In the top panel we compare our models to observational data from \cite{Hashimoto18} at $z = 9$, \cite{Harikane20} at $z \sim 6$, \cite{Hashimoto19} at $z \sim 7$, \cite{Inoue16} at $z \sim 7$, \cite{Witstok22} at $z = 6.7 - 7$, \cite{Fujimoto22} at $z = 8.5$, \cite{Algera23} at $z = 7.3$, \cite{Laporte_2017} at $z = 8.3$, \cite{Tamura19} at $z = 8.31$, \cite{Ren23} at $z = 7.2$, \cite{Carniani17} at $z \sim 7$, \cite{Popping22} (upper limit) at $z = 13$, and \cite{Schouws24,Schouws25} at $z = 14.2$. We used the relations from \cite{Harikane20} for $z = 6-9$, as well as the SB and high-redshift relations from \cite{deLooze14}. In the bottom panel we compare our models to the simulated values from \cite{Katz22} ($z = 6$), \cite{Pallottini19} ($z = 6$), \cite{Lupi2020} ($z = 6$), \cite{Pallottini22} ($z = 7.7$), and \cite{Padilla23} ($z = 6$). The relations are from the simulations of \cite{Olsen17} ($z \simeq 6$) and \cite{Arata20} ($z = 6-9$). }  
    \label{fig:OIII_vs_SFR}
\end{figure}

Figure~\ref{fig:CII_vs_SFR} shows the [CII]-SFR relations obtained by the High-z model (red) and the fiducial model by \cite{Schimek23} (purple), in comparisons to $z > 6$ observations and local analogues (top panel) and simulations (bottom panel). 
We find the High-z model is in very good agreement with the lower total [CII] luminosities from $z > 6$ observations, such as the $z = 6-9$ relation from \cite{Harikane20}. Compared to simulations, the High-z model agrees with upper theoretical estimates. 
In Fig.~\ref{fig:OIII_vs_SFR} we show the [OIII]-SFR relations obtained by the High-z model (red) and \cite{Schimek23} (purple), in comparison to observational data (top panel) and simulations (bottom panel). The High-z model predicts lower total [OIII] luminosity compared to the higher values from $z > 6$ observations at comparable SFR and is in better agreement with the high-z ($z = 0.5-6.6$) observations from \cite{deLooze14}. Most of the observational [OIII] constraints in the top panel exhibit higher SFR than \textsc{Ponos}. The [OIII]-SFR prediction from the High-z model is, however, in good agreement with the theoretical predictions (bottom panel), where most of the data points lie below the observed values in the top panel.

\onecolumn
\begin{multicols}{2} 

\section{Dwarfs model and Halo model emission maps}\label{app:dwarfs}

\begin{figure*}[h!]
    \centering
    \includegraphics[width=0.7\textwidth]{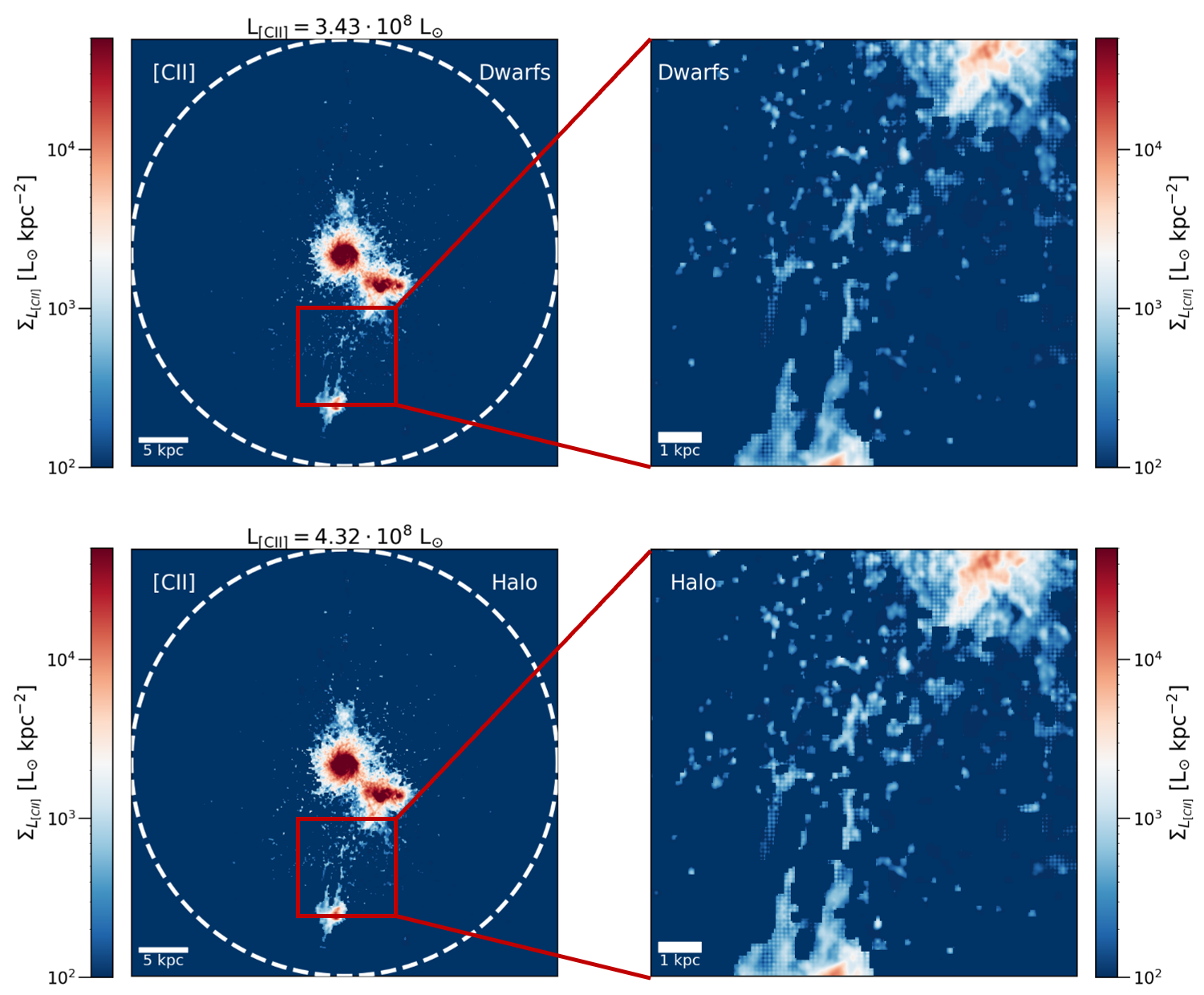}
    \caption{Emission maps of the $\mathrm{[CII]}$ line obtained with the Dwarfs model (top panels) and the Halo model (bottom panels). The right panels are zoomed-in images if the tidal tail between merger B and merger A. The dashed circles mark the extent of the virial radius of \textsc{Ponos}.}  
    \label{fig:CII_maps2}
\end{figure*}

In this section, we show the [CII] and [OIII] line emission maps and [OIII]/[CII] ratio maps obtained with the Dwarfs and Halo models. The $\mathrm{[CII]}$ maps are shown in Fig.~\ref{fig:CII_maps2}, with the Dwarfs model maps in the top panels and the Halo model maps in the bottom panels. The white dashed circles mark the virial radius of \textsc{Ponos} ($R_{\rm{vir}} = 21.18$ kpc), and the right panels zoom in on the tidal tail between the merger B and merger A components. The Dwarfs model maps show a slightly weaker surface brightness in the disc and merger B component compared to the Solar model map, which reflects the overall lower ($18\%$) total [CII] luminosity . The Halo model maps appear almost identical to the Solar model maps, which reflects the very small increase ($3\%$) in total [CII] luminosity. 

\begin{figure*}[ht]
    \centering
    \includegraphics[width=0.7\textwidth]{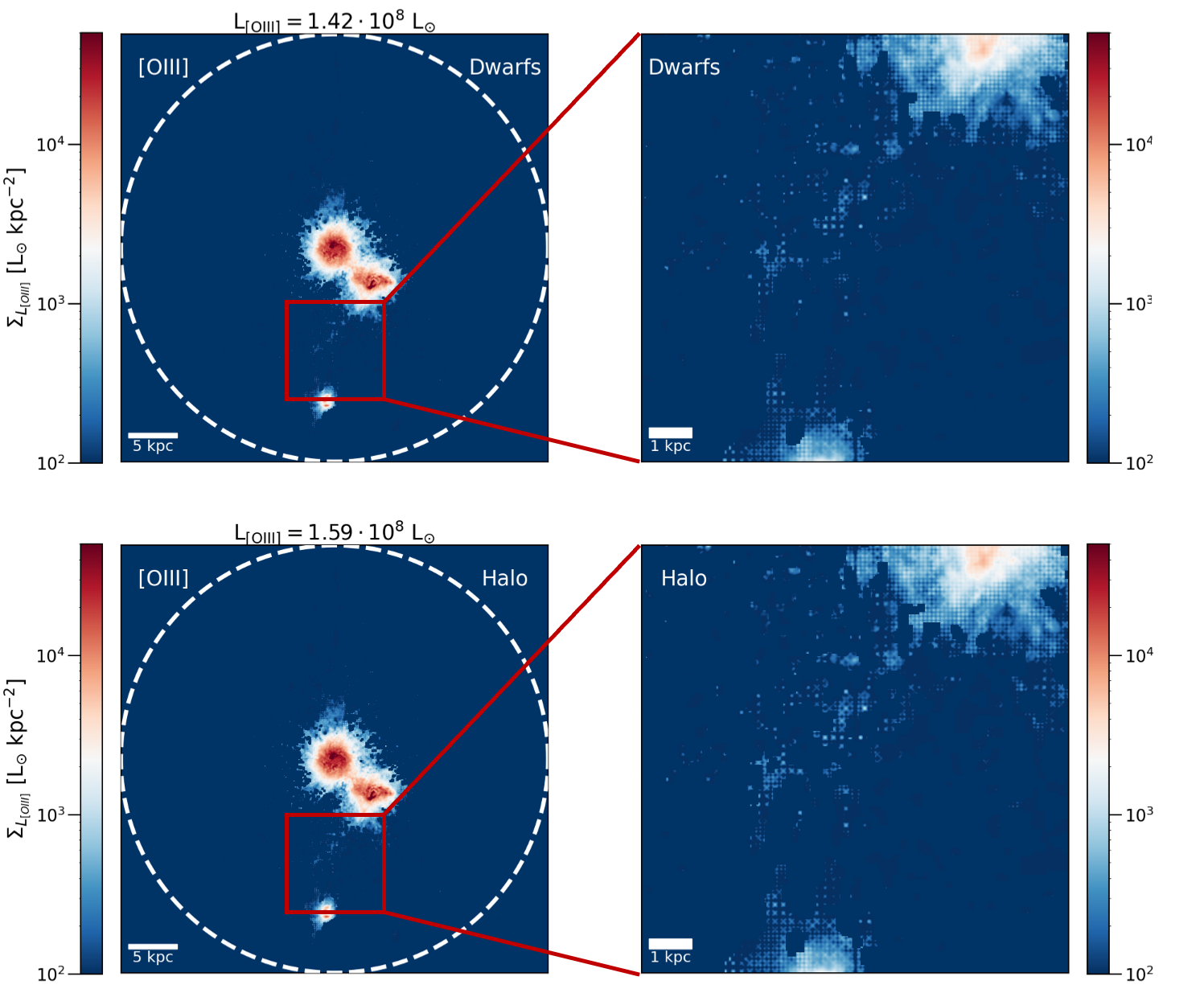}
    \caption{Emission maps of the $\mathrm{[OIII]}$ line obtained with the Dwarfs model (top panels) and the Halo model (bottom panels). The right panels are zoomed-in images of the tidal tail between merger B and merger A. The dashed circle marks the extent of the virial radius of \textsc{Ponos}.}  
    \label{fig:OIII_maps2}
\end{figure*}

In Fig.~\ref{fig:OIII_maps2} we show the [OIII] emission maps obtained by the Dwarfs model (top panels) and Halo model (bottom panels), with the right panels zooming in on the tidal tail between the merger B and merger A components. The Dwarfs model maps are very similar to the High-z model maps, with stronger surface brightness in the CGM compared to the Solar model, reflecting the increased total [OIII] luminosity. The Halo model maps show a stronger surface brightness in the extended emission compared to the Solar model, where the total [OIII] luminosity obtained by the Halo model is the highest amongst our models. \\

\begin{figure*}[ht]
    \centering
    \includegraphics[width=0.7\textwidth]{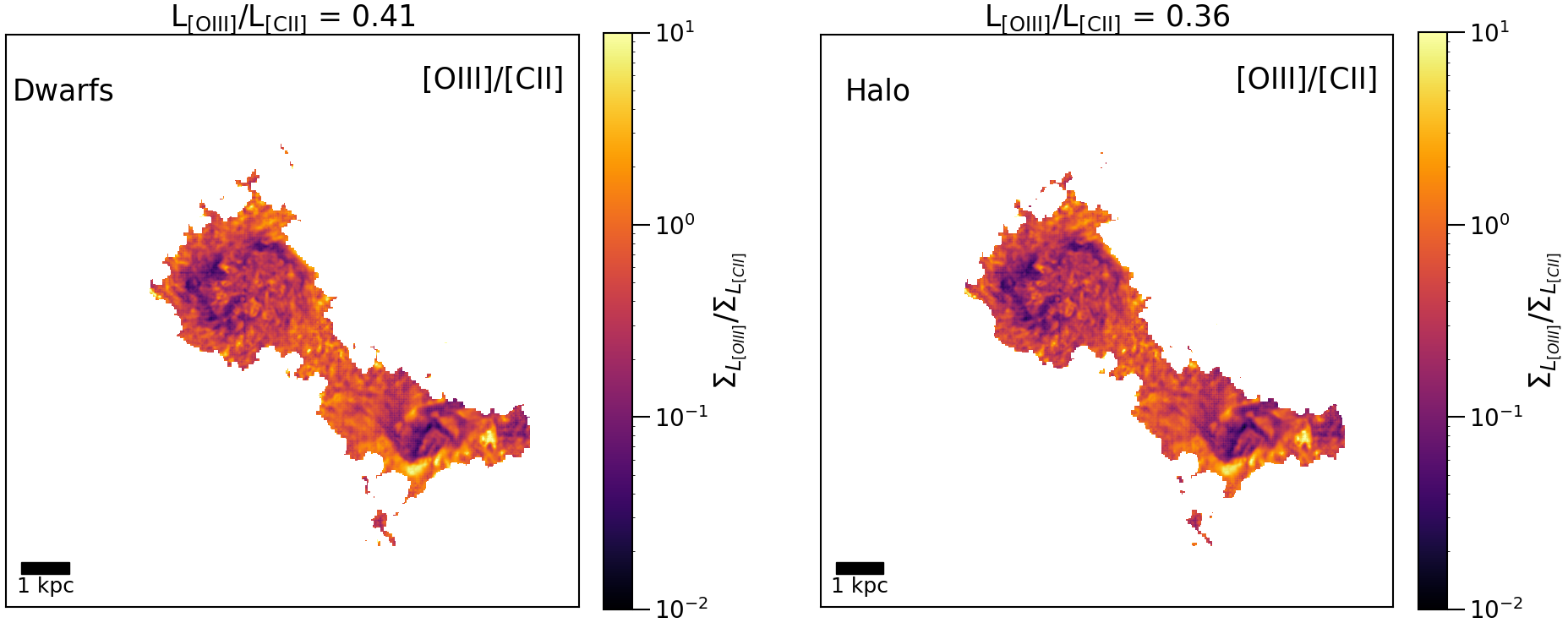}
    \caption{Ratio maps of the $\mathrm{[OIII]/[CII]}$ luminosity ratio, comparing the Dwarfs model (left) and the Halo model (right). The total luminosity ratios obtained with each model is shown in the title.}  
    \label{fig:ratio2}
\end{figure*}

Following the same method as for the Solar model and High-z model [OIII]/[CII] ratio maps (see Sect.~\ref{sec:ratio_maps}), we show the [OIII]/[CII] ratio maps for the Dwarfs model (left panel) and the Halo model (right panel) in Fig.~\ref{fig:ratio2}. Both the Dwarfs model and Halo model maps reflect the models' higher [OIII]/[CII] luminosity ratios compared to the Solar model.
\end{multicols}

\end{appendix}

\end{document}